\documentclass[twocolumn,showpacs,aps,prx,floatfix,superscriptaddress]{revtex4-1}
\usepackage[english]{babel}
\usepackage[utf8]{inputenc}
\usepackage{graphicx}
\usepackage{dcolumn}
\usepackage{epsfig}
\usepackage{amssymb}
\usepackage{amsmath}
\usepackage{natbib}
\usepackage{bm}
\usepackage{textcomp}

\renewcommand{\vec}[1]{\mathbf{#1}}

\renewcommand{\Im}{\mathop{\mathrm{Im}}}
\newcommand{\sign}{\mathop{\mathrm{sign}}}
\newcommand{\ep}{\epsilon}

\begin{document}

\title{Multiple magneto-phonon resonances in graphene}

\author{D. M. Basko}
\email{denis.basko@lpmmc.cnrs.fr}
\affiliation{LPMMC, Universit\'e de Grenoble-Alpes and CNRS,
25 rue des Martyrs, 38042 Grenoble, France}

\author{P. Leszczynski}
\affiliation{LNCMI, UPR 3228, CNRS-UJF-UPS-INSA, 38042 Grenoble,
France}

\author{C. Faugeras} \email{clement.faugeras@lncmi.cnrs.fr}\affiliation{LNCMI, UPR 3228, CNRS-UJF-UPS-INSA, 38042 Grenoble, France}

\author{J. Binder}
\affiliation{Institute of Experimental Physics, Faculty of
Physics, University of Warsaw, Poland.}

\author{A.A.L. Nicolet}
\affiliation{LNCMI, UPR 3228, CNRS-UJF-UPS-INSA, 38042 Grenoble,
France}

\author{P. Kossacki}
\affiliation{LNCMI, UPR 3228, CNRS-UJF-UPS-INSA, 38042 Grenoble,
France} \affiliation{Institute of Experimental Physics, Faculty of
Physics, University of Warsaw, Poland.}

\author{M. Orlita}
\affiliation{LNCMI, UPR 3228, CNRS-UJF-UPS-INSA, 38042 Grenoble,
France}

\author{M. Potemski}
\affiliation{LNCMI, UPR 3228, CNRS-UJF-UPS-INSA, 38042 Grenoble,
France}

\date{\today }

\begin{abstract}

Our low-temperature magneto-Raman scattering measurements performed on graphene-like locations on the surface of bulk graphite reveal a new series of magneto-phonon resonances involving both K-point and $\Gamma$ point phonons. In particular, we observe for the first time the resonant splitting of three crossing excitation branches. We give a detailed theoretical analysis of these new resonances.
Our results highlight the role of combined excitations and the importance of multi-phonon processes (from both K and $\Gamma$ points) for the relaxation of hot carriers in graphene.
\end{abstract}

\maketitle

\section{Introduction}

Quantization of electronic motion by a strong magnetic field leads to many
qualitatively new effects. They are most pronounced in two-dimensional
systems, where the motion along the field is frozen by quantum confinement,
so the continuous electronic spectrum is transformed into a series of
discrete Landau levels (LLs).
Probably the most famous examples of such new effects are the integer and
the fractional quantum Hall effects~\cite{Yoshioka2010}, which arise
because the degenerate electronic states on a LL are profoundly modified
by disorder or lateral confinement potential, and by electron-electron
interaction, respectively. Electron-phonon coupling can also produce new
effects, especially when the energy of an inter-LL excitation matches that
of an optical phonon. Such resonances may lead to strong-coupling effects
even if the electron-phonon coupling in the absence of the magnetic field
was weak.

Magneto-phonon resonances were first discussed theoretically in relation
to the dc electron transport in doped bulk semiconductors subject to a
strong magnetic field where the resonance leads to an enhancement of the
electron-phonon scattering rate~\cite{Gurevich1961}. They were later
observed experimentally in InSb~\cite{Puri1963,Shalyt1964}.
A doublet structure was observed in the infrared absorption spectrum 
of bulk InSb in a magnetic field, which was attributed to the resonant coupling of an electronic excitation, the cyclotron resonance mode, to optical phonons, when tuned in resonance~\cite{Johnson1966}. This is a strong-coupling
effect, whose detailed theory was presented in Ref.~\cite{Korovin1968}. 
Later, it was suggested that such magneto-phonon resonance need not be
restricted to doublets: three branches of excitations involving zero,
one, and two phonons may also cross at the same value of the magnetic
field, resulting in a triplet structure of the absorption
peak~\cite{Korovin1971}. In fact, equidistancy of LLs for electrons
with parabolic bands leads to a rich variety of possible resonant
combinations~\cite{Lang2005}. Still, we are not aware of any 
observation of a resonant splitting of more than two excitation branches
in conventional semiconductors.

Graphene, with its conical electron dispersion and exceptional crystal
quality, opens new possibilities for observation of fundamental
physical effects. 
In a magnetic field~$B$, perpendicular to the crystal plane, the LL
energies are $E_{\pm{n}}=\pm\sqrt{2n}(\hbar{v}/l_B)$, where $n=0,1,\ldots$
is the Landau level index, $v\approx{10}^6\:\mbox{m/s}$ is the Dirac
velocity, and $l_B=\sqrt{\hbar/(eB)}$ is the magnetic length. Due to
the $\sqrt{n}$ dependence, the spacing between LLs with small $n$ is
larger than for electrons with parabolic spectrum, which enabled the
observation of the quantum Hall effect at room
temperature~\cite{Novoselov2007}, unprecedented in conventional
semiconductor structures.
Magnetophonon resonances have been observed in Raman scattering on
optical phonons~\cite{Faugeras2009,Yan2010}, where the Raman G~peak
acquired a doublet structure. Here, we report an observation of
magnetophonon resonances where the triple peak structure is clearly
seen. To the best of our knowledge, it is the first observation of
such resonances in solid-state physics.

Despite the generality of the observed magnetophonon resonance effects,
graphene introduces some qualitatively new aspects. First, electronic
bands in graphene are conical rather than parabolic, so the LLs are
not equally spaced in energy. This leads to a classification
of possible resonant combinations which is quite different from
that in conventional semiconductors. Such classification for graphene
is developed below.
Second, while in an undoped semiconductor all LLs in the conduction
band are empty, and those in the valence band are completely filled, in
graphene there is always a partially filled LL (unless the magnetic
field is tuned to some special values); if this partially filled level
is
involved in the resonance, the physics of the electron-phonon coupling
is changed significantly. Namely, it turns out that the multiple-peak
structure cannot be viewed as a resonant splitting of a few discrete
levels; one has to face a true many-body problem, so the peak splitting
is necessarily accompanied by a broadening of a similar magnitude, and
complicated spectral shapes may be produced. This physics was overlooked
in the previous theoretical studies of magneto-phonon resonances in
graphene~\cite{Ando2007,Goerbig2007,Pound2012}.

Our experiments are performed on extremely pure flakes of graphene
which can be found on the surface of bulk graphite
\cite{Li2009,Neugebauer2009,Yan2010,Faugeras2011,Kuhne2012,Qiu2013}.
To probe the magnetophonon resonances, we use Raman scattering, which
is a powerful and popular tool for characterization of graphene
\cite{Ferrari2013}. Inter-LL electronic excitations can be observed in
the Raman spectrum of graphene~\cite{Faugeras2011,Berciaud2014}, and the
strongest features are due to transitions $-n\to{n}$~\cite{Kashuba2009}.
In addition, transitions $-n\to{n}\pm{1}$ are also
observed~\cite{Kuhne2012} even though theory predicts them to be
weaker~\cite{Kashuba2009}.
In this work, we trace the evolution of inter-LL excitations in graphene
on graphite in magnetic fields up to $B=30$~T. Such fields are needed to
tune the inter-LL excitation energies across the energies of optical
phonons and their combinations. We observe different series of
magnetophonon resonances, involving both phonons from the vicinity of the
$K$ and $\Gamma$ points of the first Brillouin zone.

The paper is organized as follows. In Sec.~\ref{sec:qualitative} we
discuss qualitatively different types of magnetophonon resonances which
can occur in graphene, and summarize our main results. In
Sec.~\ref{sec:experiment} we give a detailed presentation of our
experimental results. In Sec.~\ref{sec:theory} the theory is developed for
the various cases discussed in Sec.~\ref{sec:qualitative}.

\section{Qualitative discussion: classification of magneto-phonon
resonances and summary of main results}
\label{sec:qualitative}

\begin{figure}
\begin{center}
\vspace{0cm}
\includegraphics[width=8cm]{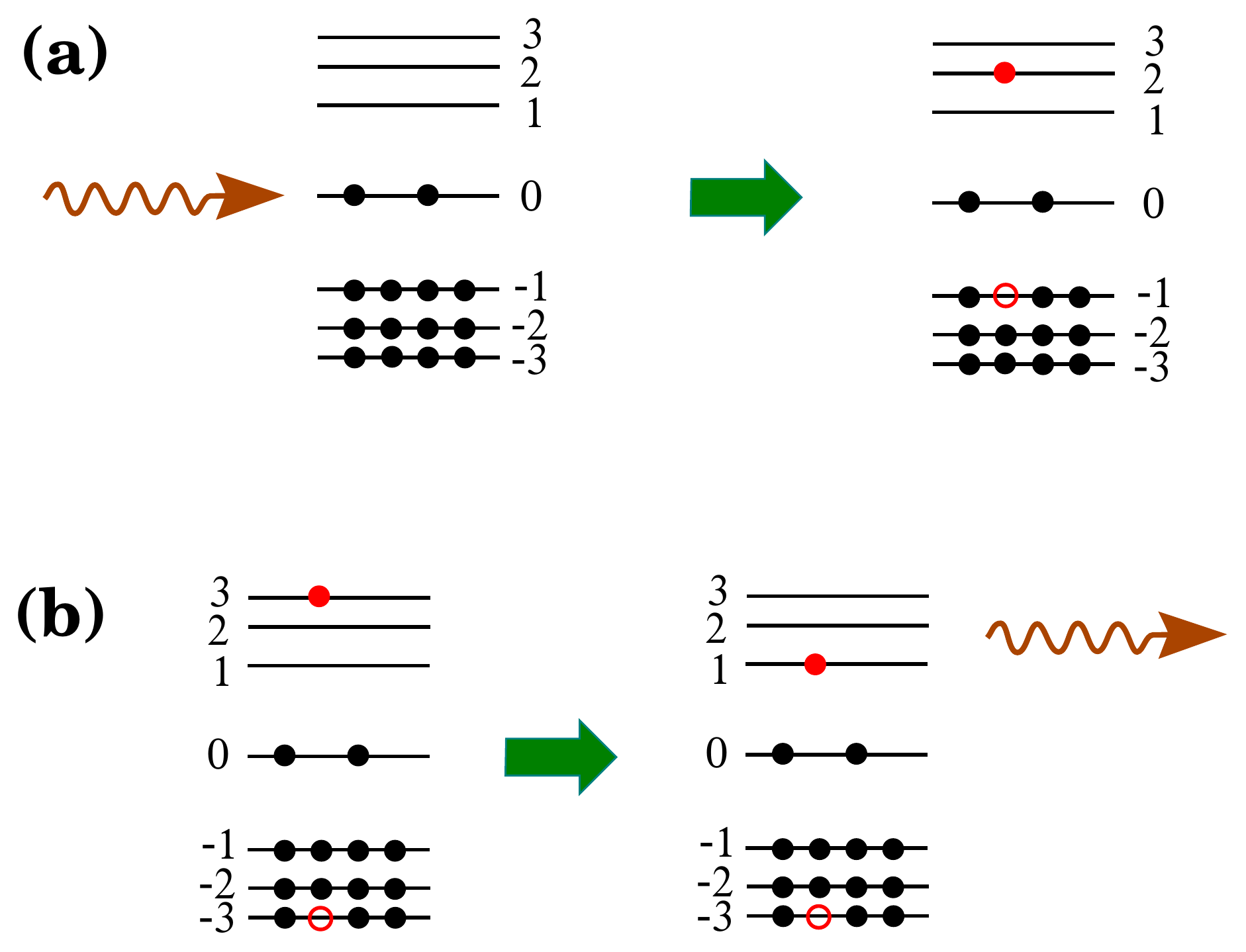}
\vspace{-0cm}
\end{center}
\caption{\label{fig:decay} (color online)
Different types of magneto-phonon resonance in excitation decay.
(a)~An optical phonon decaying into an electronic excitation.
(b)~An electronic excitation decaying into an optical phonon and
another electronic excitation.
The horizontal lines represent the Landau levels, the black
circles represent their filling. The filled red circle represents
the excited electron, the empty red circle represents the hole,
and the wavy line represents the phonon. 
}
\end{figure}

In the following, we use a shorthand notation $L_{-n,m}$ for an
excitation of an electron from a filled level $-n$ to an empty
level~$m$ (assuming that the partially filled level is $n=0$).
To gain a qualitative insight into various resonant processes, it
is instructive to analyze the weak coupling limit first. Namely,
consider an initial excitation which  was created in the sample
in the course of Raman scattering, and analyze different channels
of decay for this excitation (see Fig.~\ref{fig:decay}).

One case is when the initial excitation is a phonon; for example,
the doubly degenerate $E_{2g}$ phonon with the wave vector at the
$\Gamma$ point of the first Brillouin zone, giving rise to the
$G$~peak in the Raman spectrum. It may decay into an electronic
inter-LL excitation, as schematically shown in
Fig.~\ref{fig:decay}(a). The selection rule for the matrix element
is $n\to{n}\pm{1}$, so the phonon decay rate is strongly enhanced
when the phonon is resonant with one of the corresponding transitions,
that is, the phonon energy
$\hbar\omega_\Gamma=E_{n+1}-E_{-n}=E_n-E_{-(n+1)}$ for some~$n$.
This case corresponds to the magneto-phonon resonance in the Raman
scattering on phonons, which was observed experimentally
\cite{Faugeras2009,Yan2010}, following the theoretical prediction
in Refs.~\cite{Ando2007,Goerbig2007}.

In the present work, we focus on a different case, that of the
initial excitation being an inter-LL excitation which decays into
a phonon and another inter-LL excitation. Again, the corresponding
decay rate is enhanced at resonance, but the resonance condition
is different from the previous case. For example, if the initial
excitation is $L_{-n,n}$, it can decay if the electron on the level
$n$ or the hole on the level $-n$ emits a phonon and moves to another
level $n'$ or $-n'$, respectively ($n'<n$). The decay is resonant
if the phonon energy $\hbar\omega=E_n-E_{n'}$. It is essentially
the same process that is responsible for the magneto-phonon
resonance in electronic transport in semiconductors
\cite{Gurevich1961,Puri1963,Shalyt1964}, as well as in
magneto-optical absorption~\cite{Johnson1966,Korovin1968}.

The role of momentum conservation is quite different in the two
processes. As we focus here on optical excitation, the momentum
of the incident and scattered photons can be neglected, so in
both cases, the total momentum of the initial excitation is zero.
If the initial excitation is a single phonon,
it can be only the $E_{2g}$ optical phonon from the $\Gamma$~point,
and the total momentum of the inter-LL excitation into which it
decays should also be zero, which leads to the selection rule
$n\to{n}\pm{1}$, and thus restricts this case to the magneto-phonon
resonance  studied in
Refs.~\cite{Ando2007,Goerbig2007,Faugeras2009,Yan2010}.
For a multi-phonon initial excitation, one can also think about the
resonant enhancement of the phonon decay rate; however, the width
of the multi-phonon Raman peaks is dominated by other factors
than the phonon decay~\cite{Basko2008,Faugeras2010a,Venezuela2011},
so the modification of latter by the magnetic field is not observed
in the Raman spectra.

For an initial electronic inter-LL excitation of zero total
momentum, the momentum conservation allows the decay into a phonon
with an arbitrary momentum and another inter-LL excitation with the
opposite momentum. Still, the matrix
element is strongly suppressed unless the phonon momentum is within
$\sim{1}/l_B$ from the $\Gamma$~or $K,K'$~points of the first
Brillouin zone. Thus, one has to consider two kinds of resonances:
(i)~those involving the $E_{2g}$ optical phonons around the
$\Gamma$~point and intravalley electronic inter-LL excitations, and
(ii)~those involving intervalley inter-LL excitations and phonons
around the $K,K'$ points. Among the latter phonons, those most
strongly coupled to the electrons are the $A_{1g}$ transverse
optical phonons. The corresponding phonon energies are
$\hbar\omega_\Gamma=196\:\mbox{meV}$ and
$\hbar\omega_K=160\:\mbox{meV}$ (see Sec.~\ref{sec:experiment}).
Another important consequence of finite phonon momenta
($\sim{1}/l_B$) involved in the process, is the absence of any
selection rule on the LL indices. In the above example of the decay
of the $L_{-n,n}$ excitation into a phonon and the $L_{-n,n'}$, any
$n'$ is allowed (energy conservation requires $n'<n$).

The above discussion of decay of the initial excitation is valid
in the limit when the electron-phonon coupling is weak compared
to the typical broadening of the Landau levels or to the dispersion
of optical phonons or inter-LL excitations on the momentum scale
$\sim{1}/l_B$. In the opposite limiting case, the problem
corresponds to the coupling between discrete energy levels,
when the states become strongly mixed. The energies of these
hybrid excitations (magnetopolarons) exhibit an anticrossing as a
function of the magnetic field as the latter is swept through the
resonance region.

For an initial electronic inter-LL excitation, different types of
resonances are possible. A resonance occurs when
the phonon energy $\hbar\omega$ matches the energy difference
between some empty Landau levels:
\begin{equation}\label{resonance=}
E_n-E_{n'}=\hbar\omega,
\end{equation}
for some $n>n'\geqslant{0}$ (we assume to be at zero temperature
and zero doping, so $n=0$ is the only partially filled level).
Due to electron-hole symmetry, $E_{-n}=-E_n$,
Eq.~(\ref{resonance=}) automatically yields a similar condition
for two filled levels, $E_{-n'}-E_{-n}=\hbar\omega$.
Also, because of the $\sqrt{n}$ dependence of $E_n$, the LLs are
not equidistant, but still, some additional degeneracies may
arise. For example, if $E_4-E_1=\hbar\omega$, this automatically
yields $E_1-E_0=\hbar\omega$, while if $E_3-E_2=\hbar\omega$,
there are no additional degeneracies.

\begin{figure}
\begin{center}
\vspace{0cm}
\includegraphics[width=5cm]{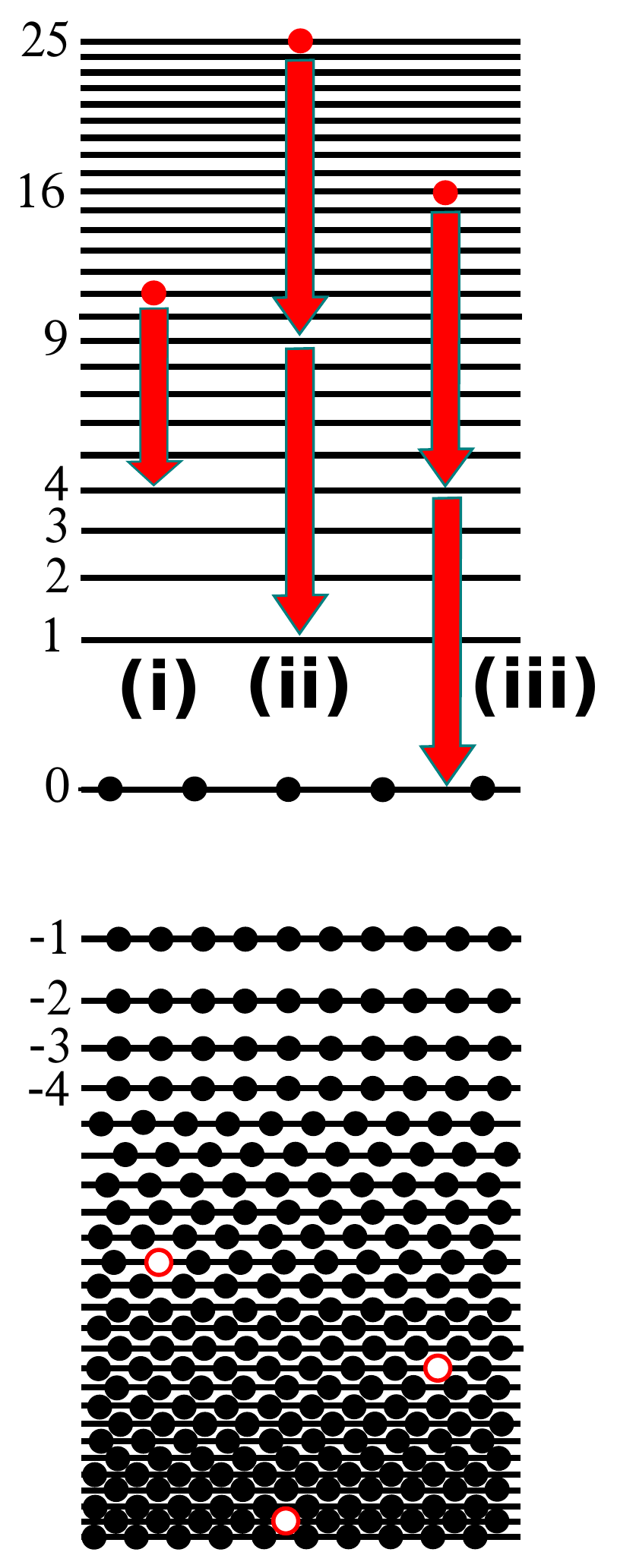}
\vspace{-0cm}
\end{center}
\caption{\label{fig:resonances} (color online)
Examples of different consequences of a resonance of the
type of Eq.~(\ref{resonance=}).
(i)~Irrational $\zeta=\sqrt{11}/(\sqrt{11}-2)$ for
the resonance $E_{11}=E_4+\hbar\omega$. No further resonant
steps are possible.
(ii)~Rational $\zeta=5/2$ for the resonance
$E_{25}=E_9+\hbar\omega$, which automatically yields the next
resonant step, $E_{25}=E_1+2\hbar\omega$. 
The $n=0$ level is not involved.
(iii)~Integer $\zeta=2$ for the resonance
$E_{16}=E_4+\hbar\omega$, which yields the next resonant step,
$E_{16}=E_0+2\hbar\omega$, involving the partially filled zero
level.
The horizontal lines represent the Landau levels,
the black circles represent their filling.
The filled red circles represent the excited electrons,
the vertical arrows show their transitions which are resonant
with a phonon. 
}
\end{figure}

All resonances of the kind (\ref{resonance=}) can be classified
into three groups, depending on whether the ratio
\begin{equation}
\zeta=\frac{E_n}{\hbar\omega}=\frac{\sqrt{n}}{\sqrt{n}-\sqrt{n'}}
\end{equation}
is integer, rational, or irrational (see Fig.~\ref{fig:resonances}).
Let us see what happens in each of these cases,
restricting our attention to the initial excitations of the type
$L_{-n,n}$, which dominate the electronic Raman scattering in the
magnetic field \cite{Kashuba2009}, and to $L_{-(n+1),n}$ (as well
as its electron-hole-symmetric counterpart $L_{-n,n+1}$), which
are also observed in the Raman spectra, as discussed in
Sec.~\ref{sec:experiment}.

(i)~If $\zeta$ is irrational, the resonances do not proliferate.
Then the simplest case is that of the initial $L_{-(n+1),n}$
excitation which is resonant with $L_{-(n+1),n'}$ plus one
phonon. In this case, there are no hole resonances; indeed,
$E_{-n''}-E_{-(n+1)}=\hbar\omega$ at the same time would imply
$\sqrt{n+1}-\sqrt{n}=\sqrt{n''}-\sqrt{n'}$ for some $n'<n''<n$,
which is impossible due to the concavity of the square root.
The symmetric excitation $L_{-n,n}$ is then resonant
with $L_{-n,n'}$ (or $L_{-n',n}$) plus one phonon, but also
with $L_{-n',n'}$ plus two phonons, i.~e. the electron and the
hole resonances necessarily occur simultaneously.
As the excitations with
zero, one, and two phonons disperse differently with the
magnetic field, the symmetric excitation gives a triple
anticrossing of the levels as a function of the magnetic field.
Such triple anticrossings were discussed long ago in the context
of optical magnetoabsorption in bulk semiconductors
\cite{Korovin1971}.

(ii)~Let $\zeta$ be rational but not an integer, and let
$m$~be the integer part of~$\zeta$. In this case,
condition~(\ref{resonance=}) yields not just one, but
$m$~resonances. Indeed, for any $j\leqslant{m}$ the energy
$E_n-j\hbar\omega$ matches the energy $E_{n_j}$ of the level
$n_j=[j\sqrt{n'}-(j-1)\sqrt{n}]^2$.
The last level in this sequence has the energy
$E_{n_m}=(\zeta-m)\hbar\omega$, which does not allow for
emission of any more phonons.
However, if we construct
$n_{m+1}=[(m+1)\sqrt{n'}-m\sqrt{n}]^2$, then it turns out
that $\sqrt{n_m}+\sqrt{n_{m+1}}=\sqrt{n}-\sqrt{n'}$, or,
equivalently, $\hbar\omega=E_{n_m}-E_{-n_{m+1}}$. This
means that the phonon itself is resonantly hybridized
with the electronic excitation from the filled level
$-n_{m+1}$ to the empty level $n_m$ (as the phonon involved
has a finite wave vector, there are no selection rules on
$n_m,n_{m+1}$).
Thus, the asymmetric excitation $L_{-(n\pm{1}),n}$ is
resonant with $L_{-(n\pm{1}),n_j}$ plus $j$ phonons for
any $j=1,\ldots,m$, so we have an anticrossing of $m+1$
energy levels with different dispersions (the hybridization
of the phonon with $L_{-n_{m+1},n_m}$ or
$L_{-n_m,n_{m+1}}$ does not add anything, as these
electronic excitations disperse with the magnetic field
in the same way as $E_n-E_{n'}$).
The symmetric excitation $L_{-n,n}$ can resonantly emit
phonons both on the electron and hole sides, up to $2m$
phonons in total, corresponding to the anticrossing of
$2m+1$ levels.
It is important that for non-integer~$\zeta$ none of the
resonances involve the partially filled $n=0$ Landau level.

(iii)~If $\zeta=m$, an integer, then the last term of the
sequence $\{n_j\}$ from the previous case is $n_m=0$.
The resonance counting is analogous to case~(ii). Namely,
we have $m+1$~anticrossings for the asymmetric initial
excitation $L_{-(n\pm{1}),n}$ and $2m+1$ anticrossings for
$L_{-n,n}$, so that the latter can thus be entirely
transformed into $2m$~phonons by a sequence of resonant
transitions.
However, involvement of the partially filled $n=0$ level
brings about a qualitatively new feature: the possibility
of creation of an arbitrary number of zero-energy
electronic excitations on the partially filled level.
As will be discussed in Sec.~\ref{sec:theoryiii}, this
feature makes a consistent theoretical treatment of
case~(iii) quite problematic.

Our experimental results, presented in Sec.~\ref{sec:experiment},
include a clear observation of several resonances of type
(iii) with double and, for the first time, triple avoided
crossings. Near each avoided crossing, we describe our data
phenomenologically by an effective two- or three-level
model with some coupling between the resonant excitations.
We also observe some signatures of resonances of type~(i),
which are, however, close to the limits of our experimental
resolution, and they do not allow for a quantitative anlysis.
Resonances of type (ii) are quite difficult to observe,
as the smallest Landau level indices $n,n'$ which produce
non-integer rational~$\zeta$ are $n=9$, $n'=1$, leading to
$\zeta=3/2$. The corresponding magnetic field is too low,
and the Landau levels involved in the electronic excitation
are too high, to be resolved in our experiment.

The theory for all three cases is presented in
Sec.~\ref{sec:theory}. Its main task is to justify (or
disprove) the phenomenological description of each avoided
crossing by an effective few-level model. Clearly, such
description may be correct only if one can neglect the
energy dispersion of the electronic excitations and of
the phonons. The former can be neglected if the Coulomb
interaction is sufficiently weak (e.g., due to screening
by the conducting graphite substrate). As for the latter,
one should recall that the most important contribution to
the  dispersion of phonons near the $\Gamma$ or $K$ points
comes from their coupling to the Dirac electrons which
produces the Kohn anomaly~\cite{Piscanec2004,Gruneis2009b}.
In a strong magnetic field the spectrum of Dirac electrons
is quantized, so the main effect of the electron-phonon
coupling is to induce magneto-phonon resonances. The
residual part of the phonon dispersion which is due to
interaction with electrons in the $\sigma$~bands is quite
weak and can indeed be neglected, as the relevant scale of
the phonon wave vectors is quite small, $q\sim{1}/l_B$
($l_B$~is the magnetic length). 

Still, it turns out, that even if one starts from
dispersionless excitations, energy dispersion can be
eventually generated due to resonant coupling. This
effect is especially dramatic for case~(iii), where
plenty of zero-energy excitations are available on the
partially filled $n=0$ Landau level. As a result, the
splitting between the energies of the coupled excitations
is accompanied by a broadening, whose magnitude is of the
same order as the splitting. The broadened peaks have
complicated spectral shapes, not described by any simple
functional form, such as Lorentzian, Gaussian, or any
other (see Fig.~\ref{fig:spectrInt} in
Sec.~\ref{sec:theoryiii} for an example). Thus, the
phenomenological description of our experimental data by
effective two- or three-level models is very approximate.

The splitting of the $L_{0,1}$ electronic excitation,
observed when it is in resonance with the $\Gamma$~point
phonon near $B=28$~T \cite{Faugeras2009,Yan2010}, can be
ascribed to either of the two types of resonances
mentioned in the beginning of this section. Indeed, on
the one hand, it can be viewed as the decay of the initial
phonon into an electronic excitation. On the other
hand, it can also be viewed as the decay of the initial
$L_{0,1}$ electronic excitation into a phonon and a
zero-energy $L_{0,0}$ excitation, falling into case~(iii)
above. Due to such ``double-faced'' nature of this
particular resonance,
the split peaks are strongly broadened. This intrinsic
broadening is missing in the simple description of
Refs.~\cite{Ando2007,Goerbig2007}.
For the resonance between the $K$~point phonon and the
$L_{0,1}$ excitation, coupling to an external continuum of
states was introduced in Ref.~\cite{Orlita2012} to describe
the strong broadening of the peaks; here we see that the
broadening, in fact, originates from the electron-phonon
coupling itself, and there is no need to invoke any
extrinsic effects.

The broadening effects are less dramatic in the cases (i)
and (ii). As discussed above, case~(ii) necessarily involves
Landau levels with quite large $n\geqslant{9}$.
This strongly suppresses the broadening effect: the main
peaks are narrow and they contain the most of the spectral
weight (Sec.~\ref{sec:theoryii}).
In case~(i), discussed in Sec.~\ref{ssec:irrational},
for asymmetric transitions $L_{-(n\pm{1}),n}$ no broadening is
generated at all, so their resonance with $L_{-(n\pm{1}),n'}$
and a phonon can be faithfully described by an effective
two-level model. No broadening arises also for symmetric
transitions $L_{-n,n}$ resonating with $L_{-n,n'}$ and the
$K$~point phonon, so the effective three-level description
is justified. However, if $\Gamma$~point phonons are involved
in the resonance, broadening is generated due to a peculiar
phonon exchange effect: the phonons emitted by $L_{-n,0}$ and
$L_{0,n}$ excitations which constitute the initial $L_{-n,n}$
excitation, can be reabsorbed in the opposite order; this is
impossible for the $K$~point phonons as the two emitted
phonons must necessarily belong to the opposite valleys.

\section{Experiment}
\label{sec:experiment}

The low-temperature magneto-Raman scattering response, at the
\textmu m scale, has been measured using a home-made miniaturized
optical probe based on optical fibers (a 5 \textmu m core
mono-mode fiber for the excitation and a 50 \textmu m core for the
collection), lenses and optical band pass filters to clean the
laser and to reject the laser line. The excitation laser at
$\lambda=514.5$~nm is focused on the sample with a high numerical
aperture aspherical lens down to a spot of the diameter
$\sim 1$~\textmu{m}. The sample is placed on piezo stages to
allow for the spatial mapping of the Raman response of the sample.
This optical set-up
is inserted in a closed jacket, with some helium exchange gas,
which is then immersed in liquid helium and placed in the center
of resistive solenoid producing magnetic fields up to 30 T. The
unpolarized Raman scattering response of our samples has been
measured in the quasi-backscattering geometry with the magnetic
field $B$ applied perpendicularly to the graphene crystal plane.
This experimental set-up has been described in more details in
Refs.~\cite{Faugeras2011,Kossacki2012}. To find a graphene flake
on the surface of bulk graphite, we use the methodology based on
the spatial mapping of the Raman scattering response of the
surface of bulk graphite with an applied magnetic field described
in Ref.~\cite{Faugeras2014}.

\begin{figure}
\includegraphics[width=1.0\linewidth,angle=0,clip]{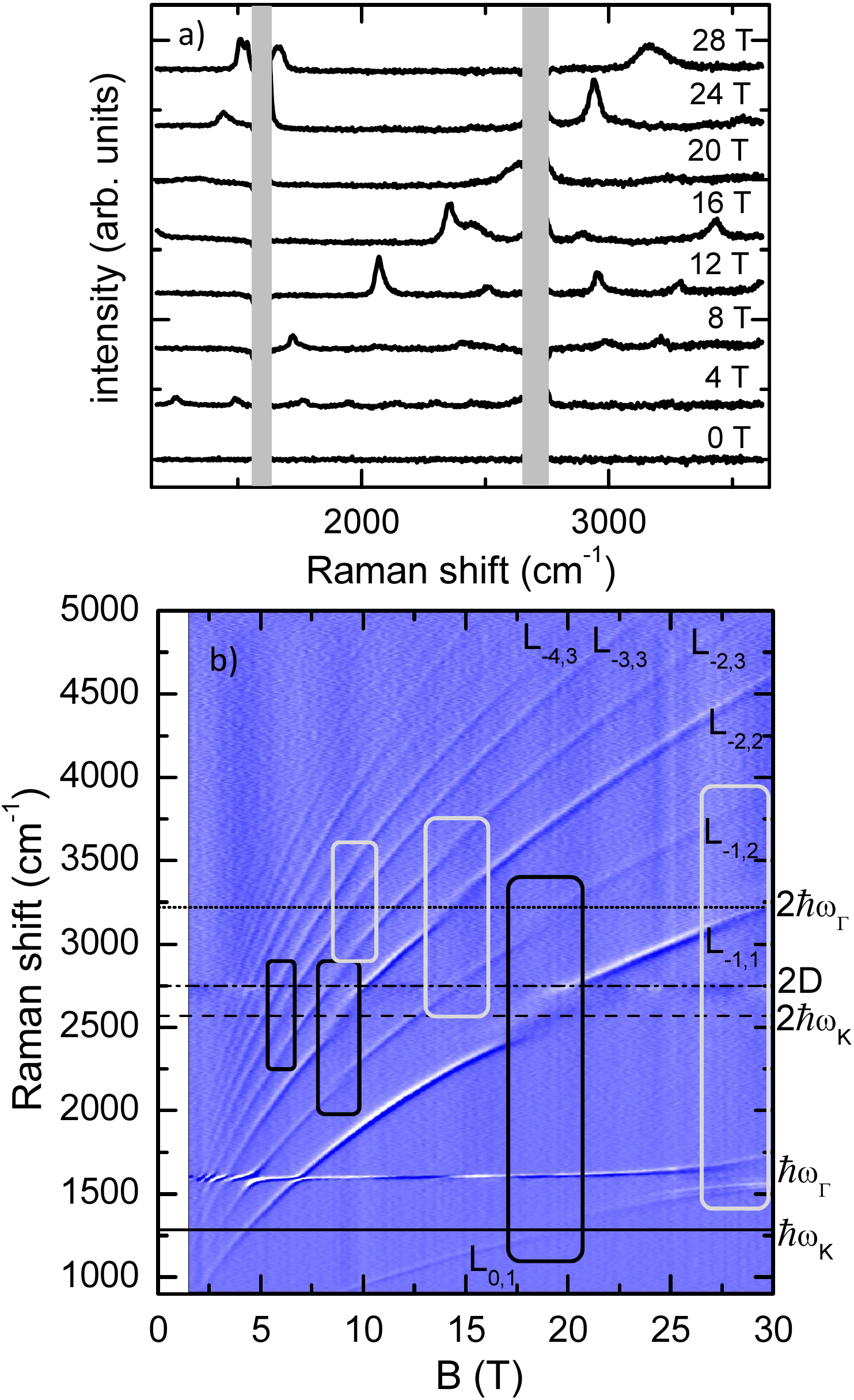}
\caption{\label{fig:spectra} 
a) Characteristic Raman scattering spectra at selected values of the magnetic field, from which the zero-field response has been subtracted. The vertical gray bars mask the residual, after normalization, phonon contributions, due to the Faraday effect in the experimental set-up. b) False color map of the Raman scattering response of the graphene-like location, differentiated with respect to~$B$, as a function of the magnetic field (with the step in $B$ of 100~mT).
Main $L_{-n,m}$ electronic excitations are indicated. The
horizontal solid line represents $\hbar\omega_{K}$, the dashed
line $2\hbar\omega_{K}$, the dashed-dotted line is at the energy
position of the 2D band feature while the dotted line represents
$2\hbar\omega_{\Gamma}$. The three white (black) boxes identify
spectral and magnetic field regions around $B=10,15$ and $28T$
($B=6,9$ and $19T$) where the electron-$\Gamma$ point (K point)
phonon resonance affects the spectra.}
\end{figure}

The typical Raman spectra, from which the $B=0$ response has been
subtracted, are shown in Fig.~\ref{fig:spectra}(a). The two gray
bars mask the two strong phonon features, the G band around
$1582$~cm$^{-1}$ and the 2D band observed at $2750$~cm$^{-1}$
for our $514.5$~nm excitation. Typically, the observed line widths
for electronic features are $\sim 30$~cm$^{-1}$. 
Due to the pronounced difference in the scattered intensity of phonon
excitations and of the different types of electronic excitations,
we present our main results in Fig.~\ref{fig:spectra}(b) in the form
of a false color map of the scattered intensity as a function of the
magnetic field which has been differentiated with respect to the
magnetic field, with a step of $100$~mT. The 2D band feature is weakly affected by the magnetic field, but because of a Faraday effect affecting the intensity of the excitation laser when sweeping the magnetic field, it cannot be cleanly removed by normalizing spectra acquired at a finite field by the $B=0$ response. Thus, differentiation as a function of $B$ provides the cleanest color map. 

The rich series of avoided crossings around $1582$~cm$^{-1}$ is
due to the well-studied magneto-phonon resonance in the Raman
scattering on the $\Gamma$-point optical 
phonon~\cite{Ando2007,Goerbig2007,Faugeras2009,Yan2010}. In the
following, instead of tracing the evolution of the phonon feature
with~$B$, we study the  evolution of the inter-LL electronic
excitations. This approach is today only possible for graphene on
graphite which shows an exceptionally high
quality~\cite{Neugebauer2009}, surpassing suspended
graphene and graphene on BN~\cite{Berciaud2014,Faugeras2015} in
terms of line widths and of the variety of the observed excitations.
The numerous features in Fig.~\ref{fig:spectra}(b) dispersing
with the magnetic field can be attributed to inter-Landau-level
electronic excitations~\cite{Faugeras2011,Kuhne2012}, and in the
present experiment they can be observed up to energies as high as
$5000$~cm$^{-1}$. We observe $L_{-n,n}$ and $L_{-(n+1),n}$ (together
with its electron-hole-symmetric counterpart $L_{-n,n+1}$) with
$n$ up to~6. Their evolution with magnetic field can be reproduced
using the value $v=(1.01\pm{0}.01)\times{1}0^6\:\mbox{m/s}$ for the
electronic velocity in graphene.

\begin{table}
\begin{tabular}{|c|c|c|}
\hline magnetic field & resonant levels & transitions affected\\
\hline 28.5~T & $E_1=\hbar\omega_\Gamma$ &
$L_{0,1}$, $L_{-1,1}$, $L_{-1,2}$\\
\hline 18.7~T & $E_1=\hbar\omega_K$ & $L_{0,1}$, $L_{-1,1}$, $L_{-1,2}$\\
\hline 14.2~T & $E_2=\hbar\omega_\Gamma$ & 
$L_{-1,2}$, $L_{-2,2}$, $L_{-2,3}$ \\
\hline 9.5~T & $E_3=\hbar\omega_\Gamma$ &
$L_{-2,3}$, $L_{-3,3}$, $L_{-3,4}$ \\
\hline 9.4~T & $E_2=\hbar\omega_K$ & $L_{-1,2}$, $L_{-2,2}$, $L_{-2,3}$ \\
\hline 6.2~T & $E_3=\hbar\omega_K$ & $L_{-2,3}$, $L_{-3,3}$, $L_{-3,4}$ \\
\hline
\end{tabular}
\caption{The values of the magnetic fields, evaluated theoretically,
and the corresponding resonances which can be observed in Fig.~\ref{fig:spectra}(b).}
\label{tab:resonances}
\end{table}

The monotonous rise of the electronic excitation energies with
the magnetic field in Fig.~\ref{fig:spectra}(b) is seen to be
interrupted at several particular values of the magnetic field,
identified by white and black boxes. These values of $B$ match
very well those corresponding to the resonances
$E_n=\hbar\omega_{K,\Gamma}$ for $n=1,2,3$,
shown in Table~\ref{tab:resonances}. In the classification of
Sec.~\ref{sec:qualitative}, all these resonances correspond to
case~(iii) with $\zeta=1$, so they are supposed to produce
double anticrossings for $L_{-(n+1),n}$, $L_{-n,n+1}$ excitations,
and triple anticrossings for $L_{-n,n}$ excitations. It is clearly
seen in Fig.~\ref{fig:spectra}(b) that the resonance of $L_{-n,n}$
excitations with the two-phonon excitation occurs at an energy
$2\hbar\omega_K\sim2570$~cm$^{-1}$. This energy is
different from that of the 2D band feature which involves phonons
away from K point and which appears at higher energies (about
2700~cm$^{-1}$ for a $780$~nm excitation). 

\begin{figure}
\includegraphics[width=0.9\linewidth,angle=0,clip]{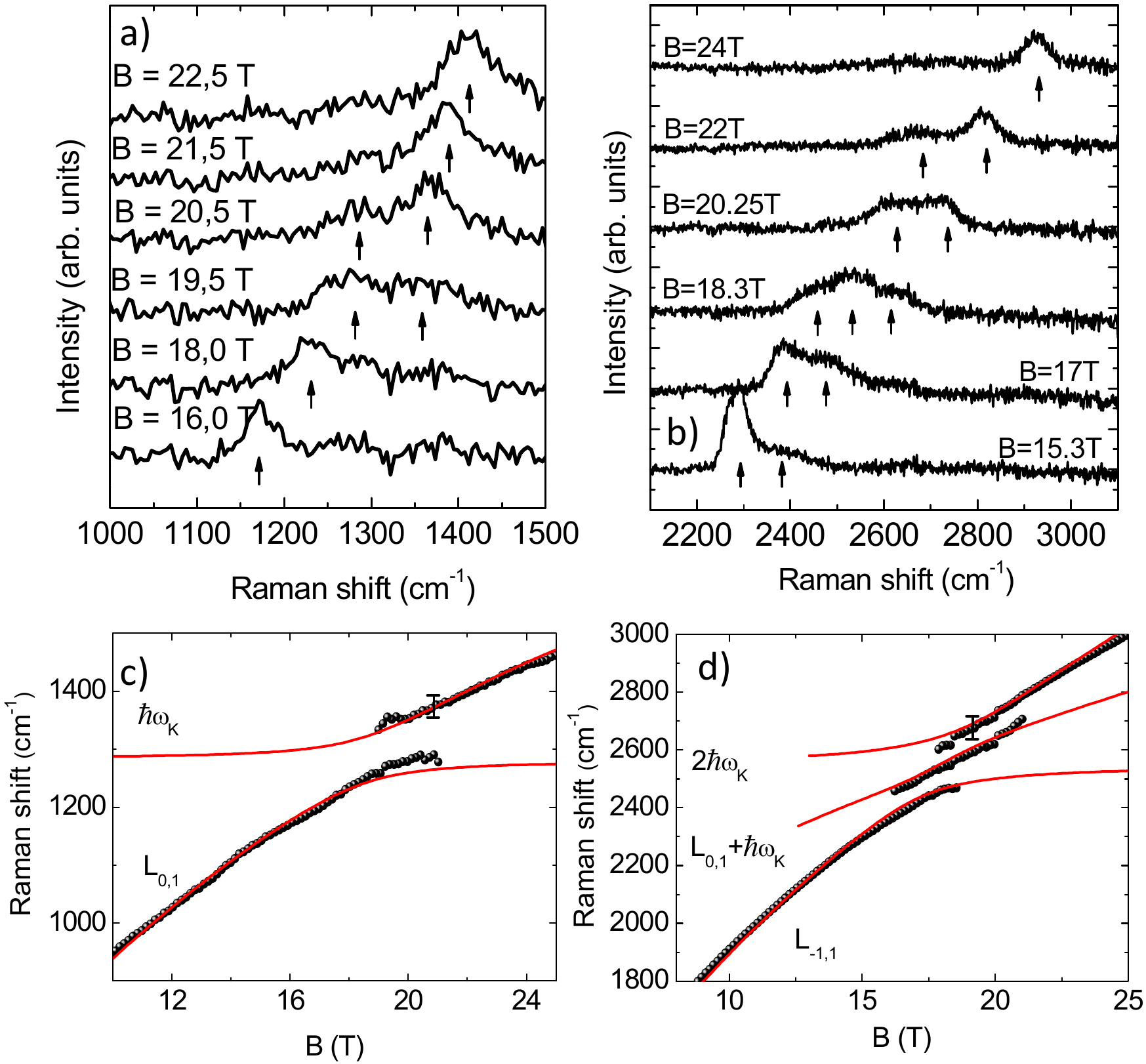}
\caption{\label{Fig3}a) Spectra of $L_{0,1}$ electronic excitation in the range of magnetic field where it is in resonance with $K$~point phonons, showing an avoided crossing behavior. b) Spectra of the $L_{-1,1}$ electronic excitation in the same range of magnetic fields as in a), and showing a triple avoided crossing. c,d) $L_{0,1}$ and $L_{-1,1}$ energies, respectively, as a function of the magnetic field (black dots) together with calculated lines (solid lines).}
\end{figure}

The difference between double anticrossings for antisymmetric
excitations and triple anticrossings for symmetric excitations
can be clearly seen in the Raman scattering spectra. In
Fig.~\ref{Fig3}(a),(b) we show the spectra corresponding to the
$L_{0,1}$ and to the $L_{-1,1}$ excitations, respectively, for selected
values of the magnetic field close to $B=18$~T. In Fig.~\ref{Fig3}(a),
the single peak observed at low magnetic field splits into two
components with a gradual transfer of its intensity to the higher
energy peak as $E_1$ crosses the phonon energy. This effect on the
$L_{0,1}$ excitation has also been observed recently in infra-red
magneto-transmission experiments performed on multi-layer epitaxial 
graphene~\cite{Orlita2012}. The evolution of the $L_{-1,1}$ 
excitation in this range of magnetic fields is completely different,
as seen in Fig.~\ref{Fig3}(b): at the resonance, a three-peak
structure appears. A similar triple avoided crossing can be observed
also on the $L_{-2,2}$ when it is tuned to $2\hbar\omega_K$ at
$B\sim9$~T, and when L$_{-1,1}$ and L$_{-2,2}$ are tuned to $2\hbar\omega_{\Gamma}$, at $B\sim28$~T and $B\sim14$~T, respectively.
Because the observed electronic excitations show a splitting, one
can conclude that we are in the strong coupling regime and that
magneto-polarons are formed at the resonant magnetic fields.

\begin{figure}
\includegraphics[width=0.9\linewidth,angle=0,clip]{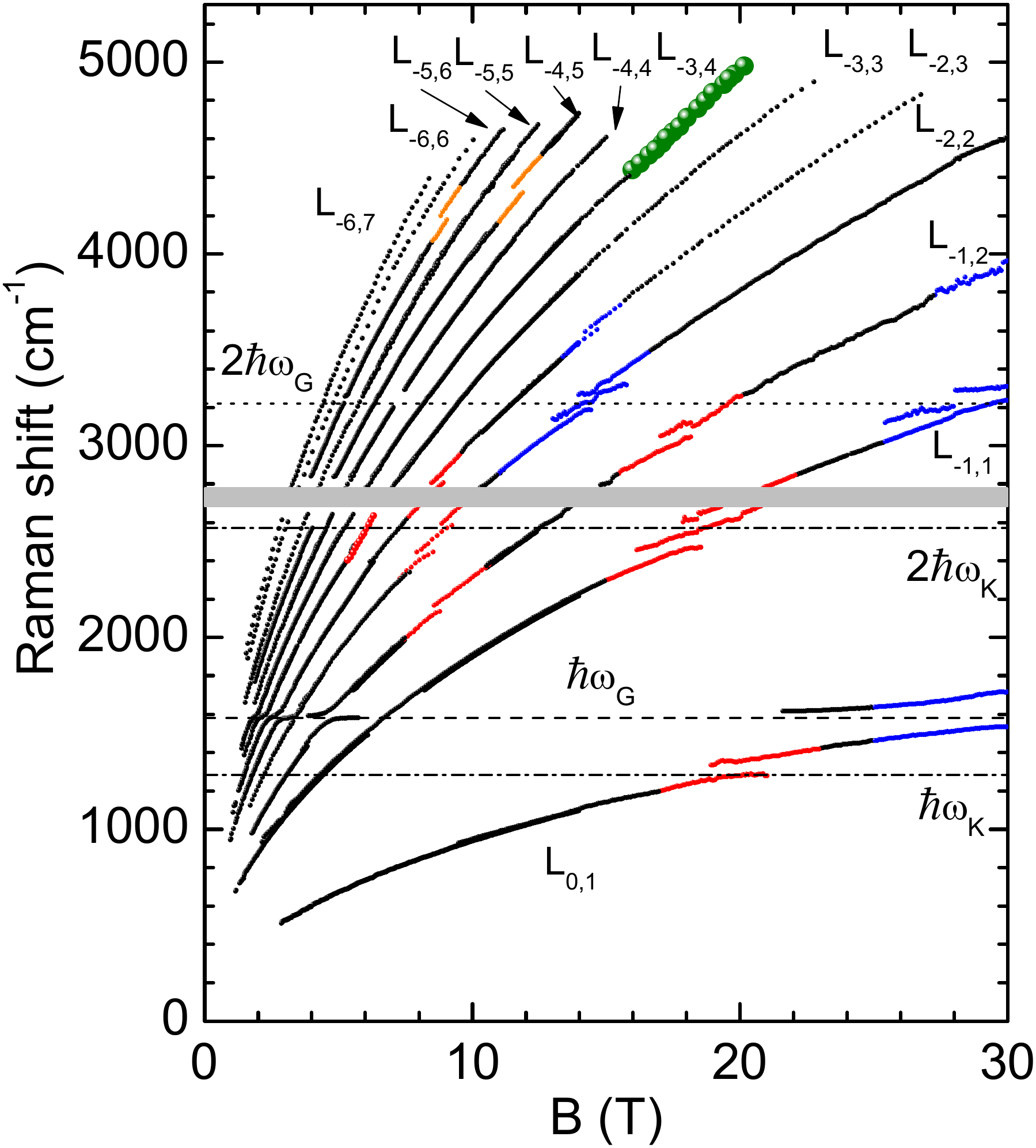}
\caption{\label{fig:anticrossings} Magnetic field dependence of the main features observed in our Raman scattering spectra. Red points indicate resonances involving K-point phonons, blue point indicate resonances involving $\Gamma$-point phonons and orange points indicate the effects of the $E_{1,n(-n,-1)}=E_K$ resonance. Green circles represent the case~(iii) resonance with $\zeta=2$ observed on the $L_{-3,4}$ excitation. The horizontal gray bar masks the residual contribution of
the 2D band.}
\end{figure}

\begin{figure}
\includegraphics[width=0.7\linewidth,angle=0,clip]{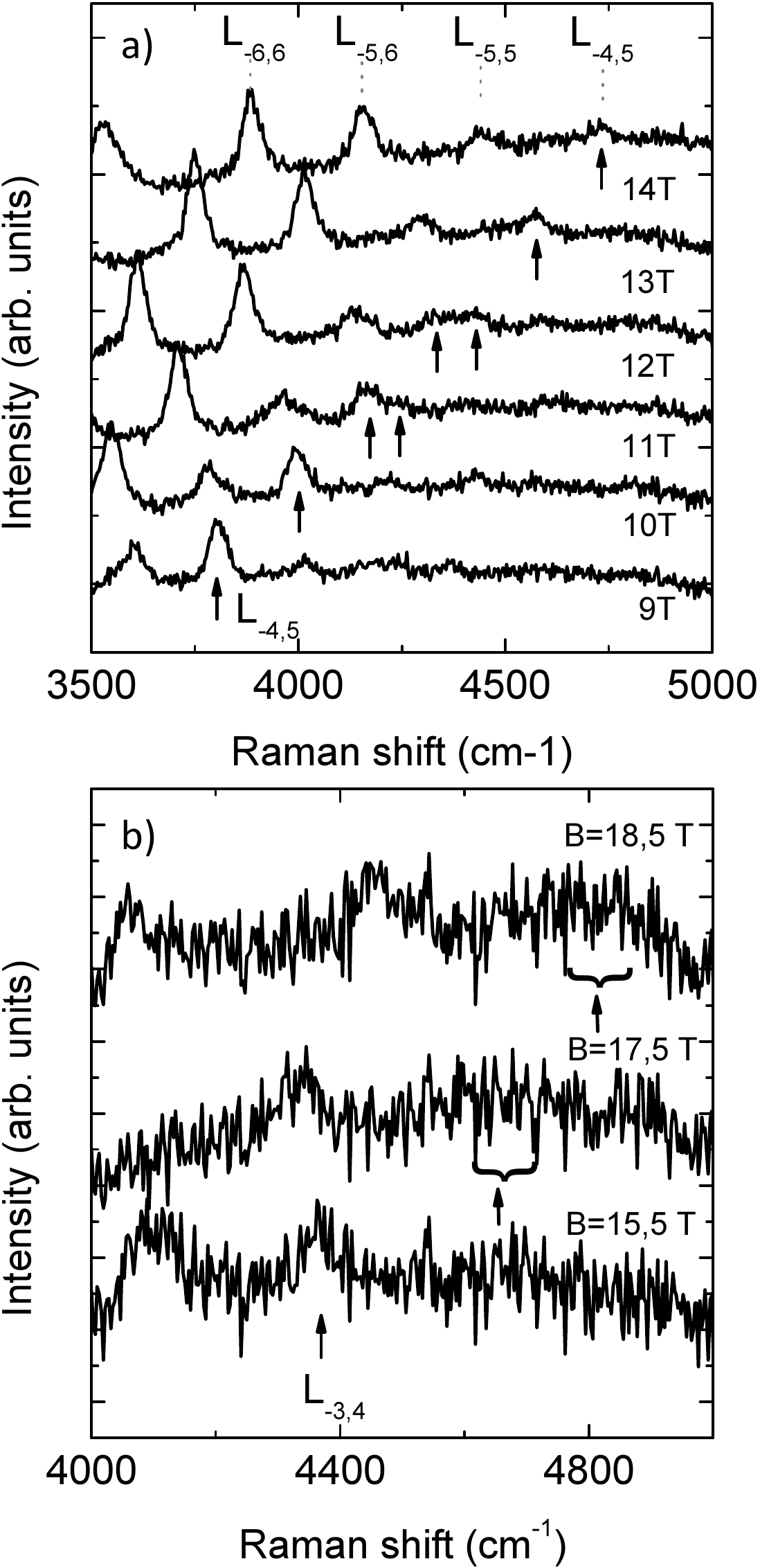}
\caption{\label{FigClass1}a) Raman scattering spectra ($B=0$~T subtracted) for selected values of the magnetic field showing the L$_{-5(-4),4(5)}$ excitation splitting around $B=12$~T. At $B=14$~T, the principal electronic excitations are identified. b) Raman scattering spectra measured at $B=15.5, 17.5$ and $18.5$~T in the range of energy of the $L_{-3,4}$ electronic excitation.}
\end{figure}

More avoided crossing features can be seen in
Fig.~\ref{fig:anticrossings}, where we present the evolution of the 
maxima of the observed electronic excitation Raman scattering 
features as a function of the magnetic field.
In this figure, in addition to the type~(iii) resonances involving
$K$-point and $\Gamma$-point phonons (red and blue points, 
respectively) mentioned in Table~\ref{tab:resonances}, some 
resonances of the type~(i) resonances involving K-point phonons can 
be seen (shown by orange points).
They correspond to the condition $E_n-E_1=\hbar\omega_K$, and are 
expected to occur at $B\approx 12.3,\, 8.9,\, 6.9$ and $5.6$~T for 
$n=5,6,7,8$, respectively. We present in Fig.~\ref{FigClass1}(a) 
characteristic Raman scattering spectra between $B=9$~T and $B=14$~T 
and showing, among others, the $L_{-4,5}$ excitation. Close to 
$B=12$~T, the corresponding Raman feature splits into two components 
and then recovers its single-peak shape. Similar behavior is observed 
near $B=9$~T on the $L_{-5,6}$ excitation. Our experimental 
resolution is not sufficient for a quantitative analysis of this 
data.
We have not been able to resolve avoided crossings at the same values
of~$B$ on $L_{-5,5}$ and $L_{-6,6}$ features. A possible reason for
this is that for $L_{-n,n}$ the splitting is necessarily accompanied 
by the intrinsic broadening, as discussed in Sec.~\ref{sec:theory}, 
which makes it harder to resolve than for asymmetric $L_{-(n-1),n}$ 
excitations. No type~(i) resonances with $\Gamma$-point phonons are 
observed either; this may be due to weaker coupling of electrons to 
these phonons than to $K$~point phonons. Finally, the feature seen at 
$B=18.7$~T on $L_{-3,4}$ line corresponds to case~(iii) with
$\zeta=2$. The Raman scattering spectra of this excitation are 
presented in Fig.~\ref{FigClass1}b) for $B=15.5, 17.5$ and $18.5$~T. The single peak observed below $B=15$~T transforms into a broad feature at high magnetic fields. This evolution is also presented in Fig.~\ref{fig:anticrossings} in the form of large green circles. Because of the lower sensitivity of our experimental setup at such high energies, we cannot fully resolve the line shape nor can we determine the number of components contributing to this feature at the resonance, although we expect that in this range of magnetic fields the resonance $E_4-E_1=E_1-E_0=\hbar\omega_K$ is relevant, so a triple avoided crossing should be observed.

It is natural to describe a double or a triple avoided crossing
phenomenologically  by an effective two- or three-level problem 
with the Hamiltonian
\begin{subequations}\begin{align}
&H_\mathrm{eff}^{2\times{2}}=\left(\begin{array}{cc}
E_n+E_{n\pm{1}} & C_1'\hbar{v}/l_B \\
C_1'\hbar{v}/l_B & E_{n\pm{1}}+\hbar\omega_\mathrm{ph}
\end{array}\right),
\label{Heff2=}\\
&H_\mathrm{eff}^{3\times{3}}=\left(\begin{array}{ccc}
2E_n & C_1\hbar{v}/l_B & 0 \\
C_1\hbar{v}/l_B & E_n+\hbar\omega_\mathrm{ph} & C_2\hbar{v}/l_B \\
0 & C_2\hbar{v}/l_B & 2\hbar\omega_\mathrm{ph} \end{array}\right).
\label{Heff3=}
\end{align}\end{subequations}
Here $\omega_\mathrm{ph}$ is either $\omega_K$ or $\omega_\Gamma$,
depending
on which resonance is considered. The coupling between different 
levels is characterized by the natural energy scale $\hbar{v}/l_B$
and the dimensionless coefficients $C_1',C_1,C_2$. For the moment,
we treat them as phenomenological fitting parameters, different for
each avoided crossing. Their values, extracted from the fit to the
experimental data are given in Table~\ref{tab:couplExpTh} for the
crossings which are sufficiently well resolved to allow for a
quantitative analysis. 

The validity of the effective models (\ref{Heff2=}), (\ref{Heff3=})
is discussed in detail in Sec.~\ref{sec:theoryiii}, where we also
give approximate relations between the phenomenological couplings
$C_1',C_1,C_2$ and the dimensionless electron-phonon coupling
constants~$\lambda_K,\lambda_\Gamma$. It is important that all
resonances involving a given phonon (at either $K$~or $\Gamma$~point)
are described by a single electron-phonon coupling constant, so the
analysis of the whole set of avoided crossings, besides producing
an estimate for $\lambda_K,\lambda_\Gamma$, provides also a
consistency check for the theory. The whole data is reasonably well
described by $\lambda_K=0.05$, $\lambda_\Gamma=0.03$, and the
comparison between theoretical and experimental values of the 
coefficients $C_1',C_1,C_2$ is given in Table~\ref{tab:couplExpTh}.


\begin{table*}
\begin{tabular}{|c|c|c|c|c|c|}
\hline transition & resonant levels & $C_1'$ or $C_1$ (exp) &
$C_1'$ or $C_1$ (th) & $C_2$ (exp) & $C_2$ (th) \\
\hline $L_{0,1}$ & $E_1=\hbar\omega_K$ & $0.044\pm 0.004$ & 
0.045 & & \\
\hline $L_{-1,1}$ & $E_1=\hbar\omega_K$ & $0.056\pm 0.003$ & 
0.063 & $0.084\pm 0.005$ & 0.070 \\
\hline $L_{-1,2}$ & $E_1=\hbar\omega_K$ & $0.042\pm 0.003$ & 0.045 & & \\
\hline $L_{-1,2}$ & $E_2=\hbar\omega_K$ & $0.049\pm 0.004$ & 0.045 & & \\
\hline $L_{-2,2}$ & $E_2=\hbar\omega_K$ & $0.055\pm 0.004$ &
0.063 & $0.088\pm 0.005$ & 0.072 \\
\hline \hline
$L_{-2,2}$ & $E_2=\hbar\omega_\Gamma$ & $0.044\pm 0.004$ &
0.049 & $0.053\pm 0.005$ & 0.057\\
\hline $L_{-2,3}$ & $E_2=\hbar\omega_\Gamma$ & $0.046\pm 0.003$ & 0.035 & & \\
\hline
\end{tabular}
\caption{The dimensionless couplings $C_1',C_1,C_2$ in the effective 
two- and three-level models (\ref{Heff2=}), (\ref{Heff3=}) for 
various avoided crossings observed in the experiment. The theoretical
values are obtained from relations given in Sec.~\ref{sec:theoryiii}
with dimensionless electron-phonon coupling constants
$\lambda_K=0.05$, $\lambda_\Gamma=0.03$.}
\label{tab:couplExpTh}
\end{table*}

\section{Theory}
\label{sec:theory}

\subsection{The model}
\label{ssec:model}

In the Landau gauge, $A_x=-By$, the electronic states in graphene
are labeled by four quantum numbers:
(i)~Landau level index~$\ell$, which is an integer running from
$-\infty$ to $+\infty$ (in the previous sections, we used
$n=|\ell|$);
(ii)~the $x$~component of the momentum~$p$, which can be taken
to run from $0$ to $L_y/l_B^2$ with spacing $2\pi/L_x$ if one
considers a finite rectangular sample of the size $L_x\times{L}_y$;
(iii)~the valley index, $K$ or $K'$;
(iv)~the spin projection, which does not play any role in the
following, so it will be omitted.
The energy depends only on the Landau level index,
\begin{equation}
E_\ell=\frac{v}{l_B}\,\sqrt{2|\ell|}\,\sign\ell,
\end{equation}
where the Dirac velocity
$v\approx{1}.0\times{10}^6\:\mbox{m/s}
\approx{6}.6\:\mbox{eV}\cdot\mbox{\AA}$,
the magnetic length
$l_B\approx(256\:\mbox{\AA})/\sqrt{B/(1\:\mbox{T})}$,
and we set $\hbar=1$ throughout this section.

We consider electron coupling to two kinds of phonons:
(i)~the $A_{1g}$ transverse optical phonons near the $K$
or $K'$ points, whose states are labeled by the valley
index $K,K'$, and the wave vector $\vec{q}$ counted from
the corresponding point ($K$~or~$K'$);
(ii)~the $E_{2g}$ optical phonons near the $\Gamma$~point,
labeled by the wave vector~$\vec{q}$ and one of the two
circular polarizations $\sigma=\pm$. We will neglect the
phonon dispersion, so the frequencies of (i)~and~(ii)
are $\omega_\Gamma\approx{196}\:\mbox{meV}$ and
$\omega_K\approx{160}\:\mbox{meV}$. It should be
emphasized that the most important contribution to the 
dispersion of phonons near the $\Gamma$ or $K$ points
comes from their coupling to the Dirac electrons which
produces the Kohn anomaly~\cite{Piscanec2004,Gruneis2009b}.
The coupling to the Dirac electrons is included in the
Hamiltonian below, so the Kohn anomaly should not be
included in the phonon dispersion to avoid double counting.
Moreover, the spectrum of Dirac electrons is strongly
modified by the magnetic field, so the resulting
contribution to the phonon dispersion is quite different
from the one at $B=0$.
Thus, what is neglected here, is the mechanical part of the
phonon dispersion which is due to the interaction with
electrons in the $\sigma$~bands. 

\begin{subequations}
The Hamiltonian is taken in the form
\begin{equation}
\hat{H}=\hat{H}_\mathrm{LL}
+\hat{H}_{\mathrm{ph},K}+\hat{H}_{\mathrm{ph},\Gamma}
+\hat{H}_{\mathrm{int},K}+\hat{H}_{\mathrm{int},\Gamma}.
\end{equation}
The non-interacting part of the Hamiltonian is given by
\begin{align}
&\hat{H}_\mathrm{LL}=\sum_{\ell=-\infty}^\infty\sum_p
E_\ell\left(\hat{c}^\dagger_{\ell,p,K}\hat{c}_{\ell,p,K}
+\hat{c}^\dagger_{\ell,p,K'}\hat{c}_{\ell,p,K'}\right),\\
&\hat{H}_{\mathrm{ph},K}=\sum_\vec{q}\omega_K
\left(\hat{b}^\dagger_{\vec{q},K}\hat{b}_{\vec{q},K}
+\hat{b}^\dagger_{\vec{q},K'}\hat{b}_{\vec{q},K'}\right),\\
&\hat{H}_{\mathrm{ph},\Gamma}=\sum_\vec{q}\omega_\Gamma
\left(\hat{b}^\dagger_{\vec{q},+}\hat{b}_{\vec{q},+}
+\hat{b}^\dagger_{\vec{q},-}\hat{b}_{\vec{q},-}\right),
\end{align}
where $\hat{c}^\dagger,\hat{c}$ are the fermionic creation
and annihilation for electrons in the corresponding
single-particle states.
The electron-phonon coupling is the same as in
Ref.~\cite{Basko2008}, but written in the Landau
level basis:
\begin{widetext}\begin{align}
&\hat{H}_{K}=\sqrt{\frac{\lambda_Kv^2}{L_xL_y}}
\sum_{\vec{q}}\sum_{\ell,\ell',p}
\tilde{J}^z_{\ell,\ell'}(-\vec{q})\,e^{ipq_yl_B^2}
\left(\hat{b}^\dagger_{\vec{q},K}
\hat{c}^\dagger_{\ell,p-q_x/2,K}\hat{c}_{\ell'\!,\,p+q_x/2,K'}
+\hat{b}^\dagger_{\vec{q},K'}
\hat{c}^\dagger_{\ell,p-q_x/2,K'}\hat{c}_{\ell'\!,\,p+q_x/2,K}\right)
+\mbox{h.c.},\label{HK=}\\
&\hat{H}_\Gamma=\sqrt{\frac{\lambda_\Gamma{v}^2}{L_xL_y}}
\sum_{\vec{q},\sigma}\sum_{\ell,\ell',p}
\tilde{J}^{-\sigma}_{\ell,\ell'}(-\vec{q})\,e^{ipq_yl_B^2}\,
\left(\hat{b}^\dagger_{\vec{q},\sigma}
\hat{c}^\dagger_{\ell,p-q_x/2,K}\hat{c}_{\ell'\!,\,p+q_x/2,K}
-\hat{b}^\dagger_{\vec{q},\sigma}
\hat{c}^\dagger_{\ell,p-q_x/2,K'}\hat{c}_{\ell'\!,\,p+q_x/2,K'}\right)
+\mbox{h.c.}.\label{HGamma=}
\end{align}\end{widetext}
where ``h.c.'' stands for the Hermitian conjugate, and the
dimensionless coupling constants $\lambda_K,\lambda_\Gamma$
are defined as in Ref.~\cite{Basko2008}. While the value
$\lambda_\Gamma\approx{0}.03$ is established relatively 
well~\cite{Pisana2007,Yan2007,Faugeras2009} and is also in
agreement with our experimental results discussed in the
previous section, the constant $\lambda_K$ is subject to
Coulomb enhancement \cite{BaskoAleiner2008,Lazzeri2008},
and thus is substrate-dependent. For our samples of graphene
on graphite, the Coulomb renormalization effects have been
shown to be strongly suppressed by substrate
screening~\cite{Faugeras2015}, so the enhancement should be
relatively weak; our experimental data are reasonably well
described by $\lambda_K=0.05$ (see the previous section).
\end{subequations}

The coefficients 
$\tilde{J}^z_{\ell,\ell'}(\vec{q})
=[\tilde{J}^z_{\ell',\ell}(-\vec{q})]^*$ and
$\tilde{J}^\pm_{\ell,\ell'}(\vec{q})
=[\tilde{J}^\mp_{\ell',\ell}(-\vec{q})]^*$,
represent the matrix elements of the phonon-induced potential
between the electronic eigenstates in the magnetic field, and
are given by
\begin{subequations}\begin{align}
\tilde{J}^z_{\ell,\ell'}(\vec{q})
={}&{}\sqrt{(1+\delta_{\ell{0}})(1+\delta_{\ell'{0}})}\,
\frac{J_{|\ell|,|\ell'|}(\vec{q})}{2}-{}\nonumber\\
&{}-(1-\delta_{\ell{0}})(1-\delta_{\ell'{0}})\sign(\ell\ell')
\frac{J_{|\ell|-1,|\ell'|-1}(\vec{q})}{2},\\
\tilde{J}^+_{\ell,\ell'}(\vec{q})={}&{}(1-\delta_{\ell{0}})
\sign\ell\,\frac{\sqrt{1+\delta_{\ell'0}}}{2}\,
J_{|\ell|-1,|\ell'|}(\vec{q}),\\
\tilde{J}^-_{\ell,\ell'}(\vec{q})={}&{}(1-\delta_{\ell'{0}})
\sign\ell'\,\frac{\sqrt{1+\delta_{\ell{0}}}}{2}\,
J_{|\ell|,|\ell'|-1}(\vec{q}).
\end{align}\end{subequations}
The formfactors,
\begin{align}
J_{n,m}(\vec{q})={}&{}(-1)^{m+\min\{n,m\}}
\sqrt{\frac{\min\{n,m\}!}{\max\{n,m\}!}}
\times{}\nonumber\\ {}&{}\times
e^{-q^2l_B^2/4}\,\left(\frac{q_x+iq_y}q\right)^{n-m}
\left(\frac{ql_B}{\sqrt{2}}\right)^{|n-m|}
\times{}\nonumber\\ {}&{}\times
\mathrm{L}_{\min\{n,m\}}^{(|n-m|)}(q^2l_B^2/2)=J_{m,n}^*(-\vec{q}),
\end{align}
are defined via the associated Laguerre polynomials,
\begin{equation}
\mathrm{L}_m^{(\alpha)}(\xi)=
\sum_{k=0}^m\frac{(-1)^k(m+\alpha)!}{k!\,(m-k)!\,(\alpha+k)!}\,\xi^k,
\end{equation}
which satisfy the following orthogonality relation:
\begin{equation}
\int\limits_0^\infty
\mathrm{L}_n^{(\alpha)}(\xi)\,\mathrm{L}_m^{(\alpha)}(\xi)\,
x^\alpha e^{-x}\,dx=\frac{(n+\alpha)!}{n!}\,\delta_{nm}.
\end{equation}

We are not interested in the Raman process itself, which was
studied in detail in Ref.~\cite{Kashuba2009}. Our main focus
is the spectral function of the electronic excitation, which
was created in this process.
The creation operators for the electronic excitations observed
in the experiment, those of the type $-n\to{n}$,
$-(n\pm{1})\to{n}$, can be represented as
\begin{equation}
\hat{R}_{n,j}^\dagger=\sum_p\hat{c}^\dagger_{n,p,K}\hat{c}_{-(n+j),p,K},
\quad j=0,\pm{1},
\end{equation}
where we have arbitrarily chosen the states in one valley, $K$,
as the states in the other valley behave exactly in the same way.
The spectral function can be obtained as
$-(1/\pi)\Im\Pi_{n,j}(\Omega)$ from the excitation propagator
\begin{equation}\label{Pinj=}
\Pi_{n,j}(\Omega)=-i\,\frac{2\pi{l}_B^2}{L_xL_y}
\int\limits_{-\infty}^\infty\left\langle\mathrm{T}\,
\hat{R}_{n,j}(t)\,\hat{R}_{n,j}^\dagger(0)\right\rangle
e^{i\Omega{t}}\,dt,
\end{equation}
where the average is over the ground state and T~stands for
the chronological time ordering.

The excitation propagator (\ref{Pinj=}) is calculated using the
standard zero-temperature diagrammatic technique~\cite{AGD},
which is constructed from the following basic elements. 
\begin{itemize}
\item
The electron Green's function, represented by a solid line, carries
the energy variable $\ep$ and the indices of the single-particle
states. It depends only on the energy and the Landau level index:
\begin{subequations}
\begin{equation}
G_\ell(\ep)=\frac{1}{\ep-E_\ell+i0^+\sign\ell}.
\end{equation}
The sign of the infinitesimal imaginary part $i0^+$ corresponds
to the Landau level filling: the levels with $\ell<0$ are filled,
those with $\ell>0$ are empty. On the level $\ell=0$ only a
fraction $f_0$ of states is assumed to filled. This can be modeled
by introducing a fictitious dependence of the infinitesimal
imaginary part on the momentum~$p$: we choose at random a fraction
$f_0$ of the momenta for which we set $G_p(\ep)=1/(\ep-i0^+)$, and
for the rest it is $G_p(\ep)=1/(\ep+i0^+)$. In the subsequent
calculations it is equivalent to simply setting
\begin{equation}
G_{\ell=0}(\ep)=\frac{1-f_0}{\ep+i0^+}+\frac{f_0}{\ep-i0^+}.
\end{equation}
\end{subequations}
\item
The phonon Green's function, represented by a wavy line, carries
the frequency variable $\omega$, the two-dimensional
momentum~$\vec{q}$, and one of the four labels $K,K',+,-$ (the
latter two correspond to the circular polarizations of the
phonons near the $\Gamma$~point). The Green's function is
given by
\begin{equation}
D_{K,\Gamma}(\omega)=\frac{1}{\omega-\omega_{K,\Gamma}+i0^+}.
\end{equation}
We do not include the negative-frequency part of the phonon
Green's function as the resonant approximation will be used in
all subsequent calculations.
\item
The electron-phonon vertices are given by the coefficients in the
interaction Hamiltonians $\hat{H}_{K,\Gamma}$ in
Eqs.~(\ref{HK=}),~(\ref{HGamma=}).
\end{itemize}

In the absence of interactions, the excitation propagator is
given by
\begin{align}
\Pi_{n,j}^{(0)}(\Omega)={}&{}\int
\frac{d\ep}{2\pi{i}}\,G_n(\ep)\,G_{-(n+j)}(\ep-\Omega)
={}\nonumber\\
={}&{}\frac{1}{\Omega-(E_n+E_{n+j})+i0^+},
\label{Pi0=}
\end{align}
so the spectral function is given by just a single
$\delta$-peak $\delta(\Omega-(E_n+E_{n+j}))$.

\subsection{Irrational case}
\label{ssec:irrational}

\begin{figure}
\begin{center}
\vspace{0cm}
\includegraphics[width=8cm]{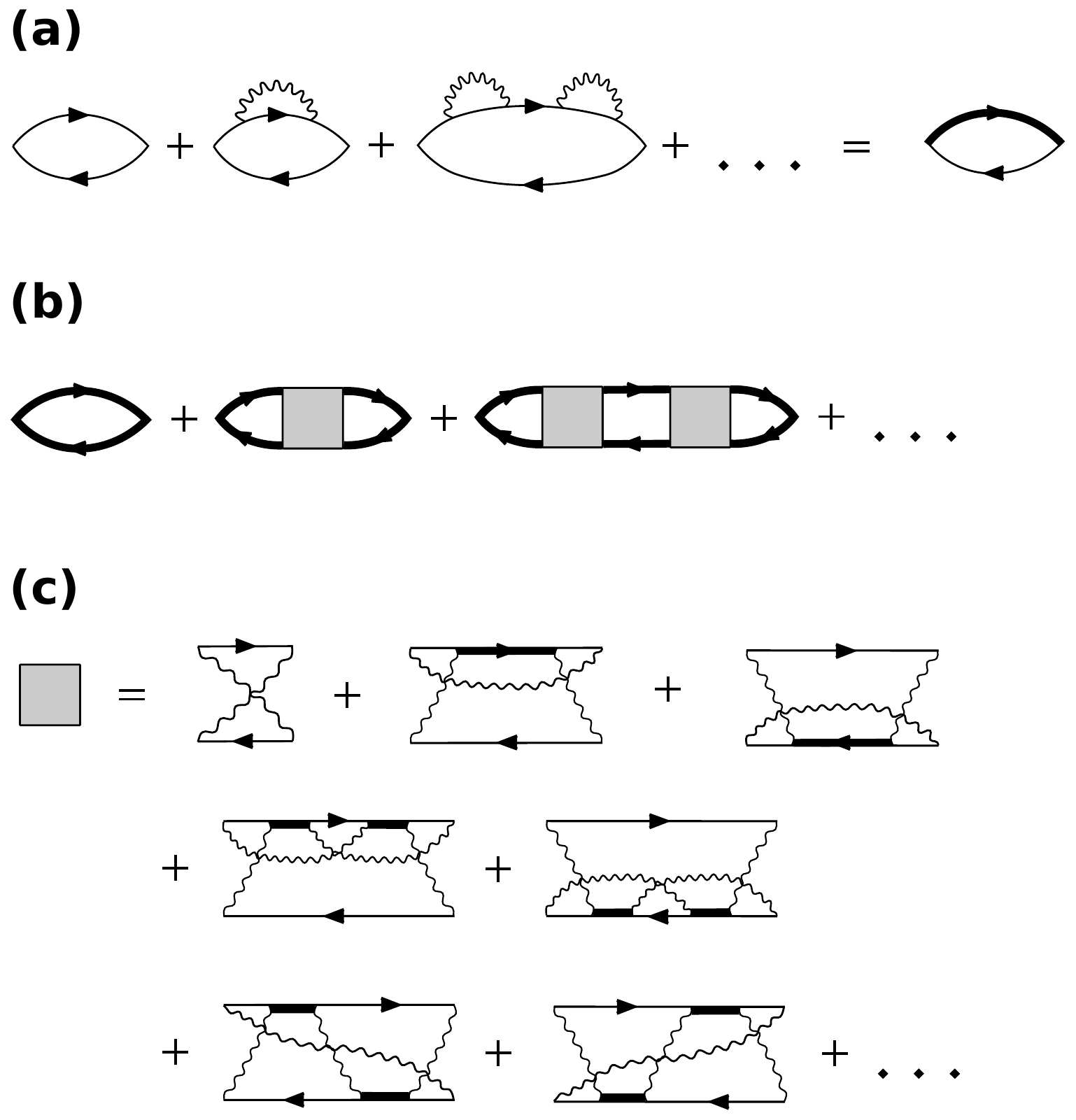}
\vspace{-0cm}
\end{center}
\caption{Diagrams contributing to the excitation propagator in
the case~(i) of Sec.~\ref{sec:qualitative},
(a)~for the asymmetric transition $L_{-(n\pm{1}),n}$,
(b)~for the symmetric transition $L_{-n,n}$, with the
vertex function given by~(c). The thick solid line represents
the dressed electronic Green's function $\tilde{G}_{\pm{n}}(\ep)$,
the shaded box represents the vertex function.}
\label{fig:irrational}
\end{figure}

Let one of the phonon frequencies $\omega_{K,\Gamma}$ be close
to the difference $E_n-E_{n'}$ and $\zeta$ is irrational
(equivalently, $\sqrt{n/n'}$ is irrational). This is case~(i)
of Sec.~\ref{sec:qualitative}. 

\subsubsection{Asymmetric transitions}

For the asymmetric
transition $-(n\pm{1})\to{n}$ the resonant contribution to
$\Pi_{n,\pm{1}}(\Omega)$ is given by the sequence of diagrams
shown in Fig.~\ref{fig:irrational}(a). It is sufficient to dress
the electronic Green's function $G_n(\ep)$ by the self-energy
\begin{subequations}\begin{align}
&\Sigma_n^K(\ep)=\lambda_Kv^2\int\frac{d^2\vec{q}}{(2\pi)^2}\,
\frac{|\tilde{J}^z_{nn'}(\vec{q})|^2}{\ep-\omega_K-E_{n'}+i0^+}
=\nonumber \\ &\qquad
=\frac{V_K^2}{\ep-\omega_K-E_{n'}+i0^+}, \label{SigmaK=}\\
&\Sigma_n^\Gamma(\ep)=\lambda_\Gamma{v}^2
\int\frac{d^2\vec{q}}{(2\pi)^2}\,
\frac{|\tilde{J}^+_{nn'}(\vec{q})|^2+
|\tilde{J}^-_{nn'}(\vec{q})|^2}{\ep-\omega_\Gamma-E_{n'}+i0^+}
=\nonumber \\ &\qquad
=\frac{V_\Gamma^2}{\ep-\omega_\Gamma-E_{n'}+i0^+}, \label{SigmaG=}\\
&V_{K,\Gamma}^2=\frac{\lambda_{K,\Gamma}v^2}{4\pi{l}_B^2},
\end{align}\end{subequations}
so that the dressed Green's function is given by
\begin{align}
\tilde{G}_{\pm{n}}(\ep)={}&{}
\frac{1}{G_{\pm{n}}^{-1}(\ep)-\Sigma_{\pm{n}}(\ep)}=\nonumber\\
={}&{}\frac{\ep\mp(\omega_{K,\Gamma}+E_{n'})}%
{[\ep\mp(\tilde{E}_n^+-i0^+)][\ep\mp(\tilde{E}_n^--i0^+)]},
\label{tildeG=}
\end{align}
where $\tilde{E}_n^\pm$ are the two roots of the equation
\begin{equation}
(\tilde{E}-E_n)(\tilde{E}-E_n'-\omega_{K,\Gamma})-V_{K,\Gamma}^2=0,
\end{equation}
representing the energies of the two hybrid (magnetopolaron)
states. The resulting excitation propagator,
\begin{align}
\Pi_{n,\pm{1}}(\Omega)={}&{}
\int\frac{d\ep}{2\pi{i}}\,\tilde{G}_n(\ep)\,G_{-(n\pm{1})}(\ep-\Omega)
=\nonumber\\ ={}&{}
\frac{\Omega-E_{n\pm{1}}-E_{n'}-\omega_{K,\Gamma}}%
{(\Omega-E_{n\pm{1}}-\tilde{E}_n^+)(\Omega-E_{n\pm{1}}-\tilde{E}_n^-)},
\end{align}
where the positive infinitesimal imaginary part of $\Omega$ is
omitted for compactness, coincides with the upper left element
of the matrix $(\Omega-H_\mathrm{eff}^{2\times{2}})^{-1}$, where
\begin{equation}
H_\mathrm{eff}^{2\times{2}}=\left(\begin{array}{cc}
E_{n\pm{1}}+E_n & V_{K,\Gamma} \\ 
V_{K,\Gamma} & E_{n\pm{1}}+E_{n'}+\omega_{K,\Gamma}
\end{array}\right)
\end{equation}
can be viewed as the Hamiltonian of an effective two-level model.
The resulting spectral function is given by the sum of two peaks
$\delta(\Omega-\tilde{E}_n^\pm)$ with the weights determined by
the corresponding eigenvectors.

\subsubsection{Symmetric transitions}
\label{sssec:irrsymm}

As discussed in Sec.~\ref{sec:qualitative}, in the case of a 
symmetric transition $-n\to{n}$, it is natural to expect the
problem to be analogous to that of three coupled levels.
However, the situation turns out to be qualitatively different
for phonons near $K,K'$ points and for those near $\Gamma$~point.

For the resonance $E_n-E_{n'}\approx\omega_K$, it is sufficient
to use the dressed Green's function both for the electron on the
level~$n$ and for the hole on the level $-n$, Eq.~(\ref{tildeG=}),
and to calculate $\Pi_{n,0}(\Omega)$ as in Eq.~(\ref{Pi0=}), but
using the dressed Green's functions instead of the bare ones.
This corresponds to the first term in the sequence of diagrams
shown in Fig.~\ref{fig:irrational}(b), and gives
\begin{align}
\Pi_{n,0}(\Omega)&{}=\int\frac{d\ep}{2\pi{i}}\,
\tilde{G}_n(\ep)\,\tilde{G}_{-n}(\ep-\Omega)={}
\nonumber\\ &{}=
\frac{(\Omega-E_n'-\tilde{E}_n^+)(\Omega-E_n'-\tilde{E}_n^-)-V_K^2}%
{(\Omega-2\tilde{E}_n^+)(\Omega-2\tilde{E}_n^-)
(\Omega-\tilde{E}_n^+-\tilde{E}_n^-)},
\label{Pinovertex=}
\end{align}
which is indeed equivalent [in the same sense as above, i.~e. 
matching the upper left matrix element of
$(\Omega-H_\mathrm{eff})^{-1}$]
to an effective three-level system with the Hamiltonian
\begin{equation}
H_\mathrm{eff}^{3\times{3}}=\left(\begin{array}{ccc}
2E_n & \sqrt{2}\,V_K & 0 \\
\sqrt{2}\,V_K & E_n+E_{n'}+\omega_K & \sqrt{2}\,V_K \\
0 & \sqrt{2}\,V_K & 2E_{n'}+2\omega_K \end{array}\right).
\end{equation}
Accordingly, the spectrum consists of three $\delta$~peaks.

For the resonance $E_n-E_{n'}\approx\omega_\Gamma$, there are
vertex corrections. This gives rise to the rest of the terms
in Fig.~\ref{fig:irrational}(b), whose contribution is of the 
same order as the first one. The vertex, in turn, is represented
by an infinite sum of diagrams, whose first terms are shown in
Fig.~\ref{fig:irrational}(c). Physically, it corresponds to the 
electron and the hole exchanging the emitted phonons before their
reabsorption. This process is possible only for the $\Gamma$~point
phonons; for the $K,K'$ phonons, those emitted by the electron and
by the whole necessarily belong to different valleys, and cannot
be exchanged, so all diagrams in Fig.~\ref{fig:irrational}(b)
vanish except for the first one.

Inclusion of vertex corrections significantly increases the level
of computational difficulty. Even though the diagrams for the
vertex function can be summed, this summation reduces the problem
to a system of integral equations for functions of momentum
$\vec{q}$ with a non-separable kernel. These equations are given
explicitly in Appendix~\ref{app:wavefunction}, where a different
but equivalent formulation of the problem in terms of
magnetopolaron wave functions is given. Solution of the resulting
equations is beyond the scope of the present paper, but the
qualitative properties of the result are worth discussing.

The dramatic effect of the phonon exchange is that the problem can
no longer be effectively described in terms of a few coupled
discrete levels. Let us note first that such description arose in
the previous case only because we neglected the phonon dispersion
as well as the dispersion of electronic excitations as a function
of the total momentum. Had we included the dispersions, already in
the simplest case of the self-energies (\ref{SigmaK=}),
(\ref{SigmaG=}), they would not reduce to a simple single-pole
form, so the spectral function $\Im\Pi_{n,\pm{1}}(\Omega)$,
instead of being a sum of two $\delta$~peaks, would broaden into a
continuum formed by excitations of different momenta. In the case
of phonon exchange, even if one originally starts from
non-dispersive excitations, the dispersion is produced as a result
of the momentum-dependent coupling between the one-phonon and the 
two-phonon sectors, so the spectral function also broadens into a
continuum with a non-trivial shape. It is important that this
broadening should be of the same order as the splitting at
resonance, as both are determined by the same energy
scale~$V_\Gamma$.

\subsubsection{Phonon dispersion}

The phonon frequencies acquire a resonant correction when
$\omega_K$ or $\omega_\Gamma$ is close to the energy
$E_n-E_{-n'}$ of a transition between filled and empty states.
For $\sqrt{n/n'}$ irrational, such resonances are distinct from
those analysed above. The shifted frequencies can be found from
the equation $\tilde{D}_{K,\Gamma}^{-1}(\vec{q},\omega)=0$,
where the dressed phonon Green's functions
$\tilde{D}_{K,\Gamma}(\vec{q},\omega)$ are given by the
diagrams shown in Fig.~\ref{fig:rational}(a).
Summation of the series gives
\begin{subequations}\begin{align}
&\tilde{D}_K^{-1}(\vec{q},\omega)
=\omega-\omega_K-4V_K^2\,\frac{|\tilde{J}^z_{n,-n'}(\vec{q})|^2
+|\tilde{J}^z_{n',-n}(\vec{q})|^2}%
{\omega-E_n-E_{n'}+i0^+},\label{tDKnnp=}\\
&\tilde{D}_\Gamma^{-1}(\vec{q},\omega)
=\omega-\omega_\Gamma-8V_\Gamma^2\,
\frac{|\tilde{J}^\sigma_{n,-n'}(\vec{q})|^2
+|\tilde{J}^\sigma_{n',-n}(\vec{q})|^2}%
{\omega-E_n-E_{n'}+i0^+}.\label{tDGnnp=}
\end{align}\end{subequations}
The coefficient is twice larger for the $\Gamma$ phonons
(8 instead of 4) because they can excite intravalley electron-hole
pairs in any valley, while a $K$~phonon can excite an electron
in $K'$ valley and a hole in $K$ valley only (the spin degeneracy
has also been taken into account).

The roots of the quadratic equation
$\tilde{D}_{K,\Gamma}^{-1}(\vec{q},\omega)=0$
for each $\vec{q}$ give two frequencies of the coupled
excitations which are superpositions of the phonon and
the electronic excitations. As $J_{nn'}(\vec{q})$ quickly
falls off for $q\gg{1}/l_B$, the resonance modifies the
phonon dispersion in the region $q\sim{1}/l_B$,
$|E_n+E_n'-\omega_{K,\Gamma}|\sim{V}_{K,\Gamma}$, and the
typical magnitude of the frequency shift in this region is
given by $V_{K,\Gamma}\sim\sqrt{\lambda_{K,\Gamma}}\,v/l_B$.
This square-root dependence on the coupling constant should
be contrasted to the Kohn-anomaly correction to the phonon
dispersion without magnetic field, which is of the order of
$\lambda_{K,\Gamma}vq$.

\subsection{Rational case}\label{sec:theoryii}

The smallest Landau level indices $n,n'$ which produce a 
non-integer rational~$\zeta$ are $n=9$, $n'=1$, leading to
$\zeta=3/2$. Even this resonance, $E_9-E_1=\omega_{K,\Gamma}$,
is beyond our experimental resolution. Still, it is not too
far from the observed resonance $E_6-E_1=\omega_{K,\Gamma}$,
so the perspective of observing the $E_9-E_1=\omega_{K,\Gamma}$
resonance in the future is not totally hopeless.
For more complicated cases, such as $E_{16}-E_1=\omega_{K,\Gamma}$
giving $\zeta=4/3$, or $E_{25}-E_9=\omega_{K,\Gamma}$ giving
$\zeta=5/2$, there is not even such hope. Thus, here we focus on
the simplest case $E_9-E_1=\omega_{K,\Gamma}$.

\begin{figure}
\begin{center}
\vspace{0cm}
\includegraphics[width=8cm]{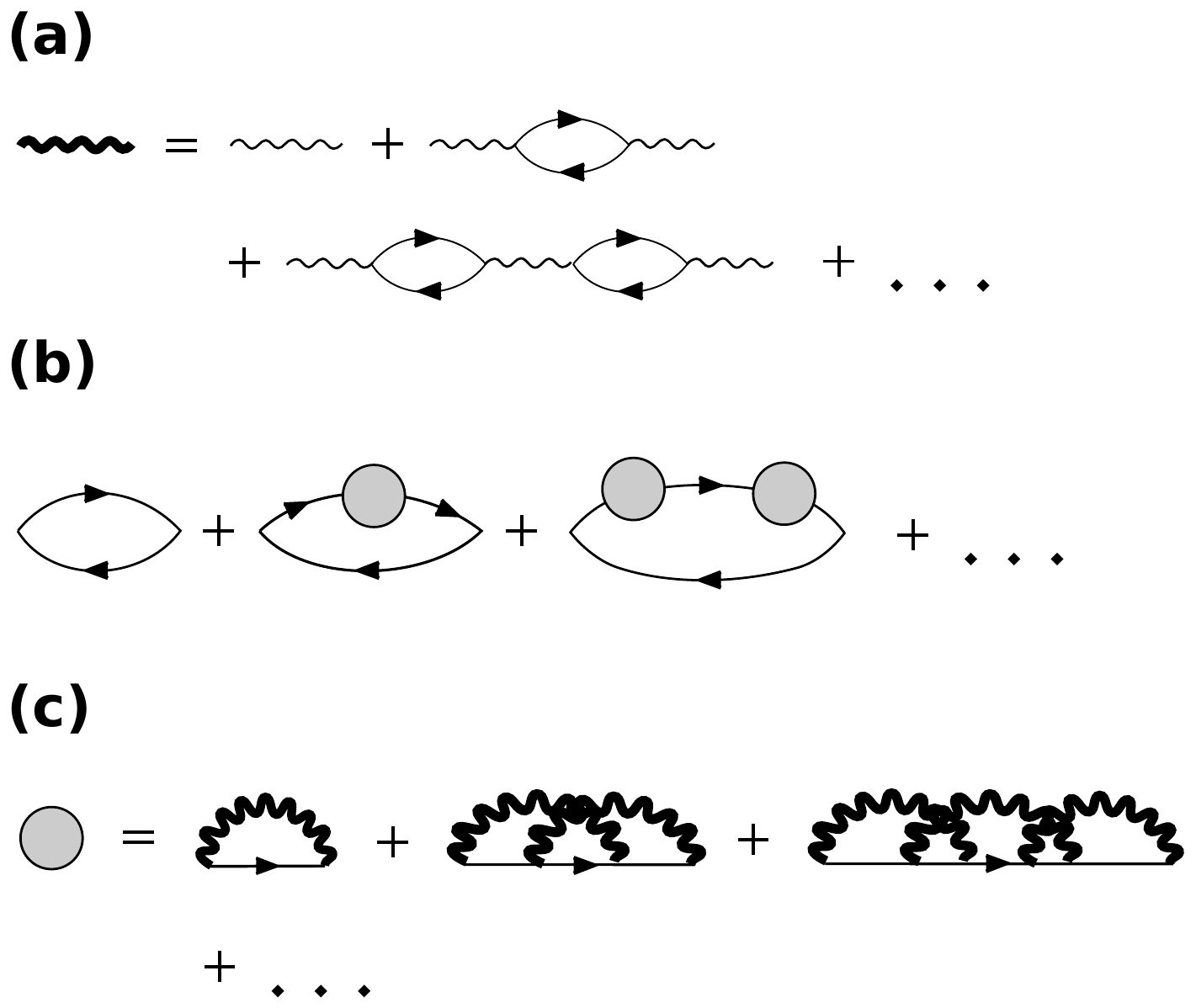}
\vspace{-0cm}
\end{center}
\caption{Diagrams contributing to the excitation propagator
$\Pi_{9,\pm{1}}(\Omega)$.
(a)~Dressed phonon Green's function.
(b)~Excitation propagator for the asymmetric transitions
$L_{-8,9}$, $L_{-10,9}$ with the electronic self-energy given
by~(c). 
The thick wavy line represents the dressed phonon Green's function
$\tilde{D}_{K,\Gamma}(\vec{q},\omega)$,
the shaded circle represents the electronic self-energy.}
\label{fig:rational}
\end{figure}

The difference of this resonance from those discussed in
Sec.~\ref{ssec:irrational} is that the condition
$E_9-E_1=\omega_{K,\Gamma}$ automatically yields
$\omega_{K,\Gamma}=E_1-E_{-1}$. Thus, the emitted phonon
can be resonantly reabsorbed by an electron on the level $n=-1$,
creating a second electron-hole pair, in addition to the one
initially produced by the photon (the energy of the resulting
excitation is the same as that of the original excitation, as
$E_9+E_{-1}=2E_1$). Thus, we dress the phonon Green's
function, as shown in Fig.~\ref{fig:rational}(a):
\begin{subequations}\begin{align}
&\tilde{D}_K(\vec{q},\omega)
=\left[\omega-\omega_K-\frac{4V_K^2|\tilde{J}^z_{1,-1}(\vec{q})|^2}%
{\omega-2E_1+i0^+}\right]^{-1},\label{tDK=}\\
&\tilde{D}_\Gamma(\vec{q},\omega)
=\left[\omega-\omega_\Gamma
-\frac{8V_\Gamma^2|\tilde{J}^\sigma_{1,-1}(\vec{q})|^2}%
{\omega-2E_1+i0^+}\right]^{-1}.\label{tDG=}
\end{align}\end{subequations}
%
For the asymmetric transitions, $L_{-8,9}$ and $L_{-10,9}$, the
excitation propagator $\Pi_{9,\pm{1}}(\Omega)$ is given by the
series in Fig.~\ref{fig:rational}(b), where the electronic Green's
function is dressed by the self-energy insertions. As in
Sec.~\ref{sssec:irrsymm}, there is a difference between the
coupling to $K,K'$ phonons and to $\Gamma$~phonons, related to
exchange. Namely, when the initial excitation (say, $L_{-8,9}$)
is resonantly converted into $L_{-8,1}$ and one $K$ or $K'$
phonon, and then into $L_{-8,1}+L_{-1,1}$, the two electrons on
the $n=1$ level must belong to different valleys, and thus are
effectively distinguishable. As a result, in the series for the
self-energy, shown in Fig.~\ref{fig:rational}(c), only the first
term survives for $K,K'$ phonons, resulting in a relatively
compact expression:
\begin{align}
\Sigma_9^K(\ep)={}&{}\lambda_Kv^2\int\frac{d^2\vec{q}}{(2\pi)^2}\,
|\tilde{J}^z_{91}(\vec{q})|^2\tilde{D}_K(\vec{q},\ep-E_1)=\nonumber\\
={}&{}\int\limits_0^\infty
\frac{V_K^2(\ep-E_9)(\xi-6)^2\xi^8e^{-\xi}\,d\xi/(2\cdot{9}!)}%
{(\ep-E_9)(\ep-\omega_K-E_1)-V_K^2(\xi-2)^2e^{-\xi}}.\label{Sigma9K=}
\end{align}
Then the excitation propagator is straightforwardly evaluated as
\begin{equation}
\Pi_{9,\pm{1}}(\Omega)=\frac{1}{\Omega-(E_9+E_{9\pm{1}})
-\Sigma_9^K(\Omega-E_{9\pm{1}})}.
\end{equation}

As discussed in Sec.~\ref{sssec:irrsymm}, as soon as excitations
involved in the decay of the initial electron-hole pair acquire
a dispersion, the peaks in the spectral function broaden and
the situation can no longer be reduced to a simple effective
picture of a few coupled discrete levels. This is what happens
here, due to the phonon dispersion acquired from the resonant
coupling to $L_{-1,1}$ electronic excitation, as seen from
Eqs.~(\ref{tDK=}), (\ref{tDG=}). However, in the present case,
this effect turns out to be strongly suppressed due to the
large Landau level index $n=9$. Let us formally expand the
self-energy $\Sigma_9^K(\ep)$ from Eq.~(\ref{Sigma9K=}) at
$V_K^2/[(\ep-E_9)(\ep-\omega_K-E_1)]\to\infty$:
\begin{align}
\Sigma_9^K(\ep)={}&{}\frac{V_K^2}{\ep-\omega_K-E_1}\times\nonumber\\
{}&{}\times\left[
1+\frac{5}{2048}\,\frac{V_K^2}{(\ep-E_9)(\ep-\omega_K-E_1)}
+\ldots\right].\label{Sigma9Kexp=}
\end{align}
At resonance, $V_K^2\sim(\ep-E_9)(\ep-\omega_K-E_1)$, the
second term in the square brackets is still small because of
the numerical factor. The origin of this numerical smallness
is that the numerator of the integrand in Eq.~(\ref{Sigma9K=})
has a maximum at $\xi\approx{12}$ and is very small at $\xi<7$,
while the function $(\xi-2)^2e^{-\xi}$ in the denominator is
quite small for $\xi>2$. Physically, this means that the
electronic transition from level $n=9$ to $n=1$ is accompanied
by emission of phonons with wave vectors $q=\sqrt{2\xi}/l_B$
which are too large and do not efficiently couple to
$L_{-1,1}$ excitation.

\begin{figure}
\begin{center}
\vspace{0cm}
\includegraphics[width=8cm]{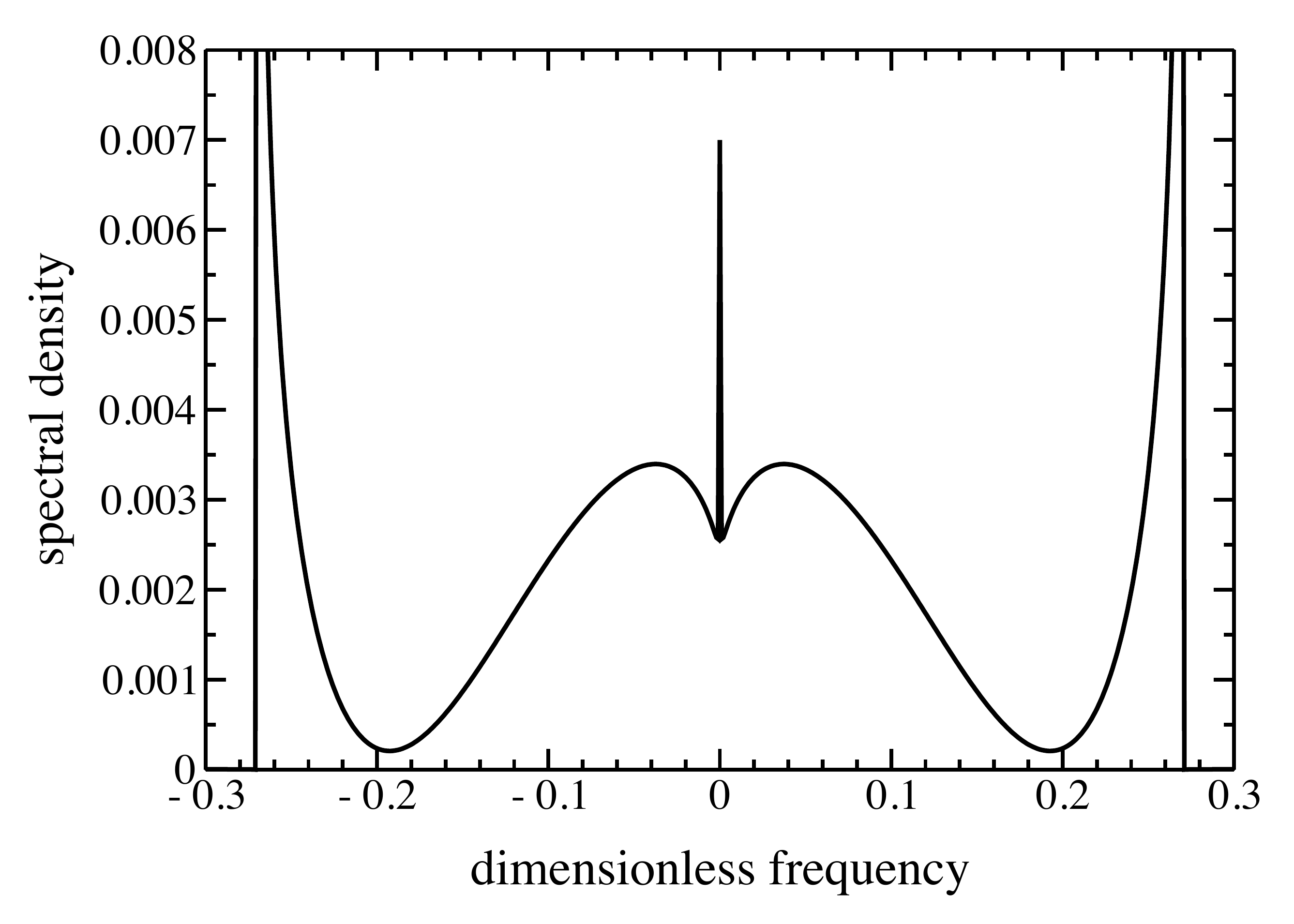}
\vspace{-0cm}
\end{center}
\caption{The dimensionless spectral density,
$-(V_K/\pi)\Im\Pi_{9,\pm{1}}(\Omega)$, versus the dimensionless
frequency $z=(\Omega-E_9/2+E_{-9\pm{1}})/V_K$ for the case of
exact resonance, $E_9-E_1=\omega_K$.
The square-root-type singularities at
$z=\pm{2}e^{-2}\approx\pm{0}.27$ correspond to the maximum of
$(\xi-2)e^{-\xi/2}$, the dimensionless energy splitting between
the phonon and the $L_{-1,1}$ excitation [see Eq.~(\ref{Sigma9K=})].
The spectral density for $|z|>2e^{-2}$ is extremely small,
$\sim{10}^{-6}$, but finite; it vanishes exactly for $|z|>2$.
The two main peaks are located near $z=\pm{1}$ (not shown).
The singularity at $z\to{0}$ is logarithmic and comes from
large~$\xi$.}
\label{fig:rationalSpectr}
\end{figure}

In the first approximation, if we neglect all terms in
Eq.~(\ref{Sigma9Kexp=}) except the first one, which is
equivalent to replacing the dressed Green's function
$\tilde{D}_K(\vec{q},\omega)$ by the bare one, i.~e.,
neglecting the coupling of the phonon to $L_{-1,1}$
excitation,
we naturally obtain the effective $2\times{2}$ problem,
\begin{equation}\label{rational2x2=}
H_\mathrm{eff}^{2\times{2}}=\left(\begin{array}{cc}
E_9 & V_K \\  V_K & E_1+\omega_K \end{array}\right),
\end{equation}
whose two eigenvalues $\tilde{E}_9^\pm$ determine the
positions of two $\delta$~peaks in the spectral function.
When the full self-energy~(\ref{Sigma9K=}) is taken into
account, the two peaks (i)~shift (the shift being equal
in magnitude and opposite in sign for the two peaks),
(ii)~broaden [as $\Im\Sigma_9^K(\tilde{E}^\pm_9)\neq{0}$,
although very small], and (iii)~lose a small part of their
spectral weight.
This part of the spectral weight is transferred to the
continuum concentrated in the vicinity of the energies
of the uncoupled excitations; in this region the spectral
function has a peculiar shape, as illustrated in
Fig.~\ref{fig:rationalSpectr} for the case of exact resonance,
$E_9-E_1=\omega_K$. At the same time, the two main peaks
have nearly Lorentzian shapes, determined by the poles
of $\Pi_{9,j}(\Omega)$ in the complex plane.
%
%
%
%
The poles are located at
\begin{align}
\Omega={}&{}E_{9+j}+\frac{E_9+E_1+\omega_K}{2}\pm\nonumber\\
{}&{}\pm\sqrt{\frac{(E_9-E_1-\omega_K)^2}{4}+V_K^2u_\pm},
\end{align}
where $u_\pm\approx 1.00251\pm 0.66\times{10}^{-6}\,i$
are the solutions of the equation
\begin{equation}
0=\mathcal{F}(u)\equiv 1-\int\limits_0^\infty
\frac{(\xi-6)^2\xi^8e^{-\xi}}{u-(\xi-2)^2e^{-\xi}}\,
\frac{d\xi}{2\cdot{9}!},
\end{equation}
and the sum of the residues at the two poles is given by
\[
\left[u\,\frac{d\mathcal{F}(u)}{du}\right]^{-1}\approx 0.9975.
\]
Thus, corrections to the simple two-level picture of
Eq.~(\ref{rational2x2=}) are quite small and will be masked
by other effects, not included in the present model.

The same can be said about the case of coupling to
$\Gamma$~point phonons. Even though the self-energy,
corresponding to the sum of diagrams in Fig.~\ref{fig:rational},
\begin{align}
\Sigma_9^\Gamma(\ep)={}&{}\sum_{k=1}^\infty
\sum_{\sigma_1\ldots\sigma_k}
\int\frac{d^2\vec{q}_1\ldots d^2\vec{q}_k}{(\pi/l_B^2)^k}\,
\tilde{J}^{\sigma_1}_{91}(\vec{q}_1)\,
[\tilde{J}^{\sigma_k}_{91}(\vec{q}_k)]^*\times{}\nonumber\\
{}&{}\times
\prod_{j=1}^{k-1}[\tilde{J}^{\sigma_j}_{1,-1}(\vec{q}_j)]^*
\tilde{J}^{\sigma_{j+1}}_{1,-1}(\vec{q}_{j+1})
\,e^{i[\vec{q}_j\times\vec{q}_{j+1}]_zl_B^2}
\times{}\nonumber\\ &{}\times
V_\Gamma^{2k}\,
\frac{\tilde{D}_K(\vec{q}_1,\ep-E_1)\ldots\tilde{D}_K(\vec{q}_k,\ep-E_1)}%
{(E_9-\ep-i0^+)^{k-1}},
\end{align}
is not easily evaluated, the factor
$\tilde{J}^\sigma_{91}(\vec{q})$ will again result in an
efficient coupling to phonons with relatively large wave
wave vectors, which disperse very little.
The same argument applies to the case of the symmetric transition
$L_{-9,9}$.
Thus, for all practical purposes, the case of rational~$\zeta$
can be considered equivalent to the irrational case.

\subsection{Integer case}
\label{sec:theoryiii}

Among integer values of $\zeta$, the most interesting one is $\zeta=1$,
that is, the resonance between the excitation $E_n-E_0$ and a phonon.
This corresponds to the most pronounced features in the experimental
data presented in Sec.~\ref{sec:experiment}. The next value, $\zeta=2$,
corresponds to the resonance $E_n-E_{n/4}=\omega_{K,\Gamma}$.
A signature of this resonance is seen as broadening of the $L_{-3,4}$
transition around $B\approx{18}\:\mbox{T}$, but the resolution is not
sufficient for a reliable quantitative analysis of the data. At the
same time, the experimental data for $\zeta=1$ resonances do allow for
a quantitative analysis.

\begin{figure}
\begin{center}
\vspace{0cm}
\includegraphics[width=8cm]{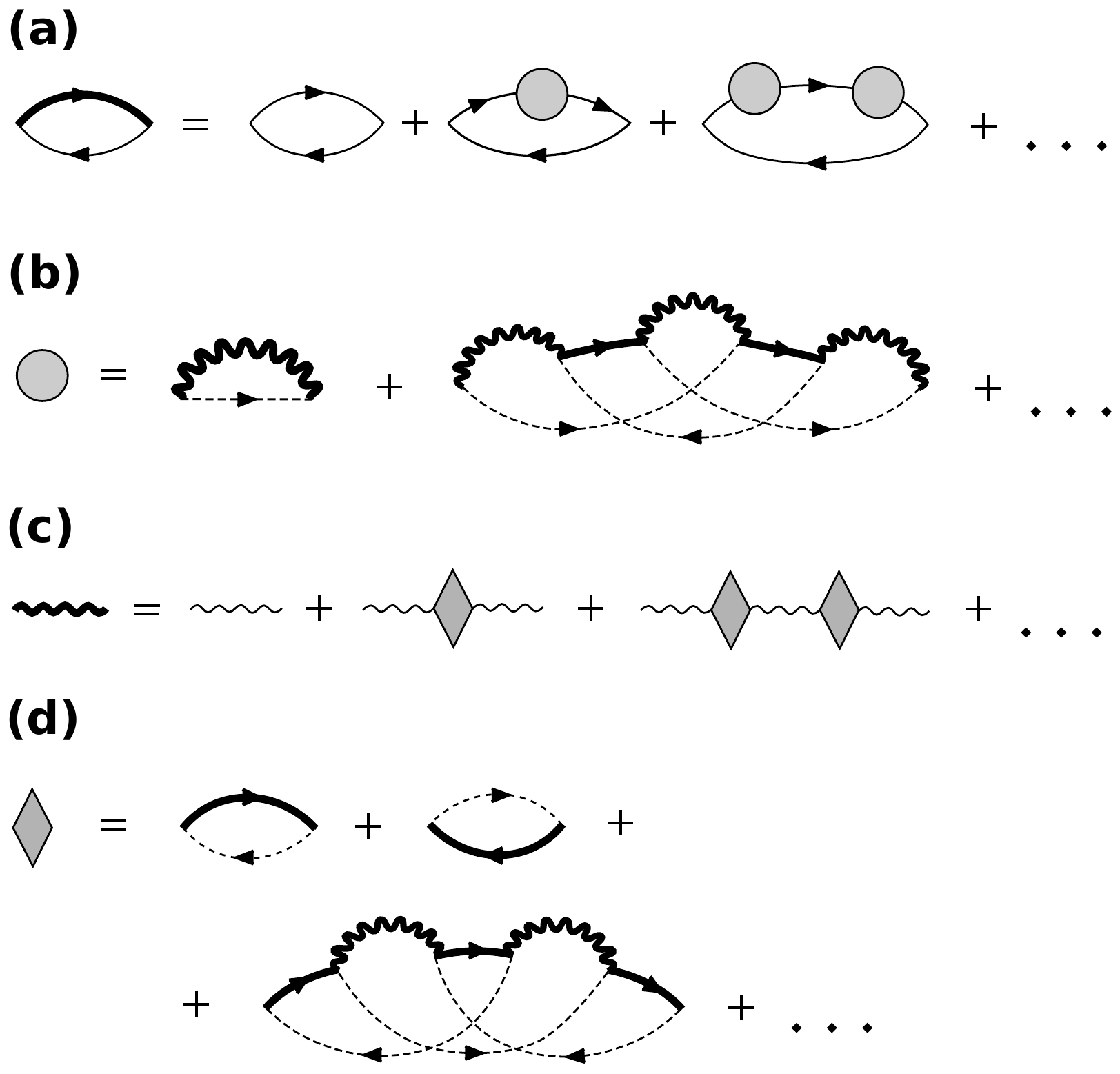}
\vspace{-0cm}
\end{center}
\caption{
(a)~Diagrams contributing to the excitation propagator
$\Pi_{n,\pm{1}}(\Omega)$ for $n>1$.
(b)~The first terms of the series for the electronic
self-energy which defines the dressed Green's functions
$\tilde{G}_{\pm{n}}(\ep)$.
(c)~Dressed phonon Green's function, defined via the
phonon polarization operator, the first diagrams for
which are shown in~(d).
Thick solid lines represent the dressed electron Green's
functions on the $n$th Landau level. Dashed lines represent
the bare electron Green's functions on the partially filled
zero Landau level. Thick wavy lines represent the dressed
phonon Green's function.
Shaded circles represent the electronic self-energy.
Shaded rhombi represent the phonon polarization operator.}
\label{fig:integerDiag}
\end{figure}

From the theoretical point of view, description of resonances
with integer~$\zeta$ represents an extremely hard problem.
In Fig.~\ref{fig:integerDiag} we show diagrams contributing
to the excitation propagator $\Pi_{n,\pm{1}}(\Omega)$ in the
simplest case of the asymmetric transition $L_{-(n\pm{1}),n}$
when $E_n-E_0$ is resonant with a phonon ($\zeta=1$) and $n>1$
[for $L_{0,1}$ transition, the diagrams for $\Pi_{1,-1}$ are
the same as in Fig.~\ref{fig:integerDiag}(c)]. The role
of electrons on the partially filled $n=0$ level is emphasized
by representing their Green's function by a dashed line.
In fact, we are unable to identify a dominant sequence of
diagrams. The physical reason for this is the macroscopic
degeneracy of the electronic non-interacting ground state due
to partial filling of the $n=0$ Landau level. As a result,
conversion of the initial electronic excitation with energy
$E_n$ into the phonon and back can be accompanied by emission
of an arbitrary number of intra-Landau-level excitations which
cost no energy and can be exchanged in arbitrary order. Thus,
various diagrams can be generated. In the
resonant region, $|E_n-\omega_{K,\Gamma}|\sim{V}_{K,\Gamma}$,
all these diagrams are of the same order, and there is no
parameter for selection of a treatable subset of diagrams.
As a result, the coupled levels broaden into a continuum, and
we are not even able to write a closed set of integral
equations to describe this broadening.

\begin{figure}
\begin{center}
\vspace{0cm}
\includegraphics[width=8cm]{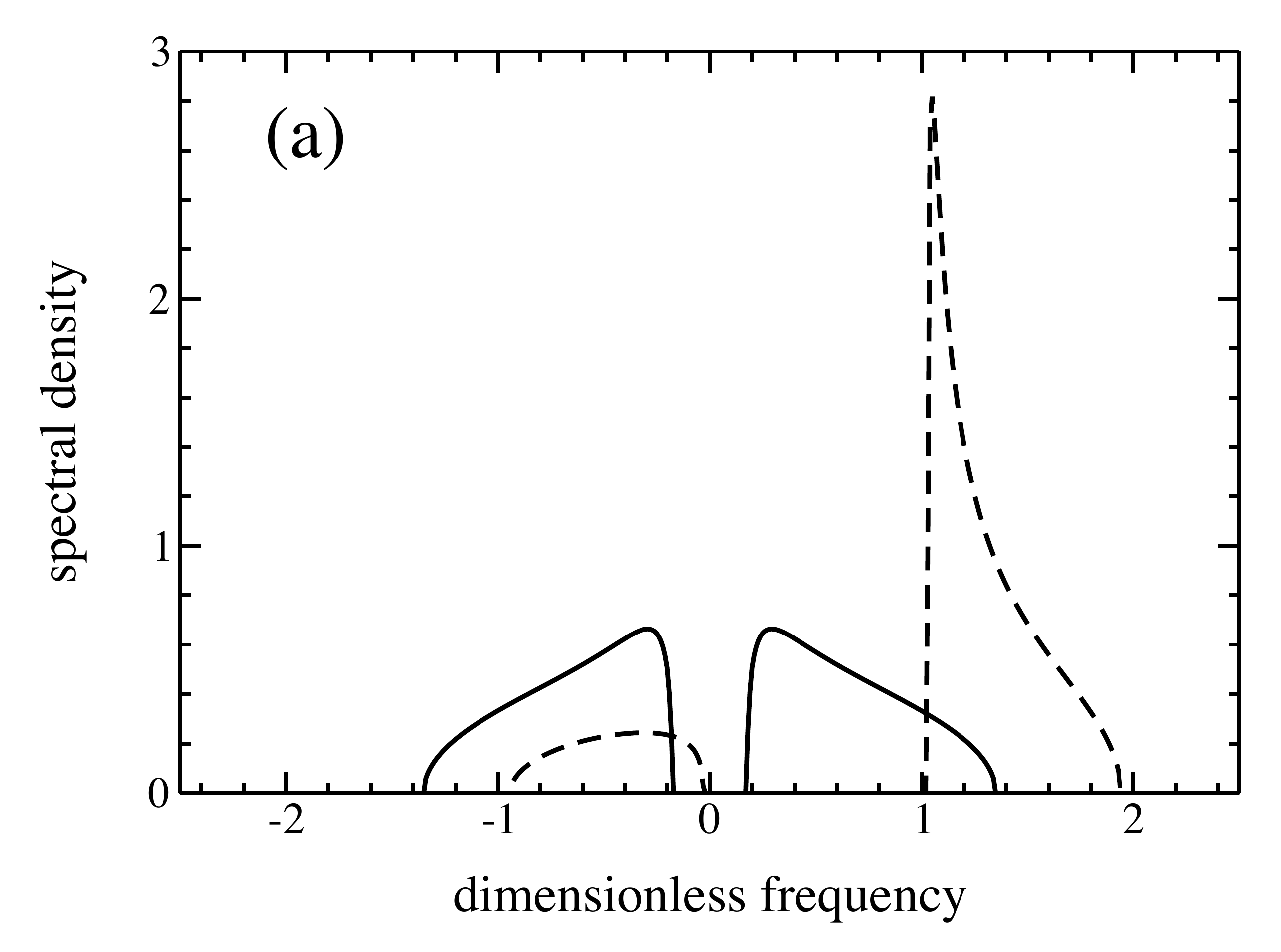}
\includegraphics[width=8cm]{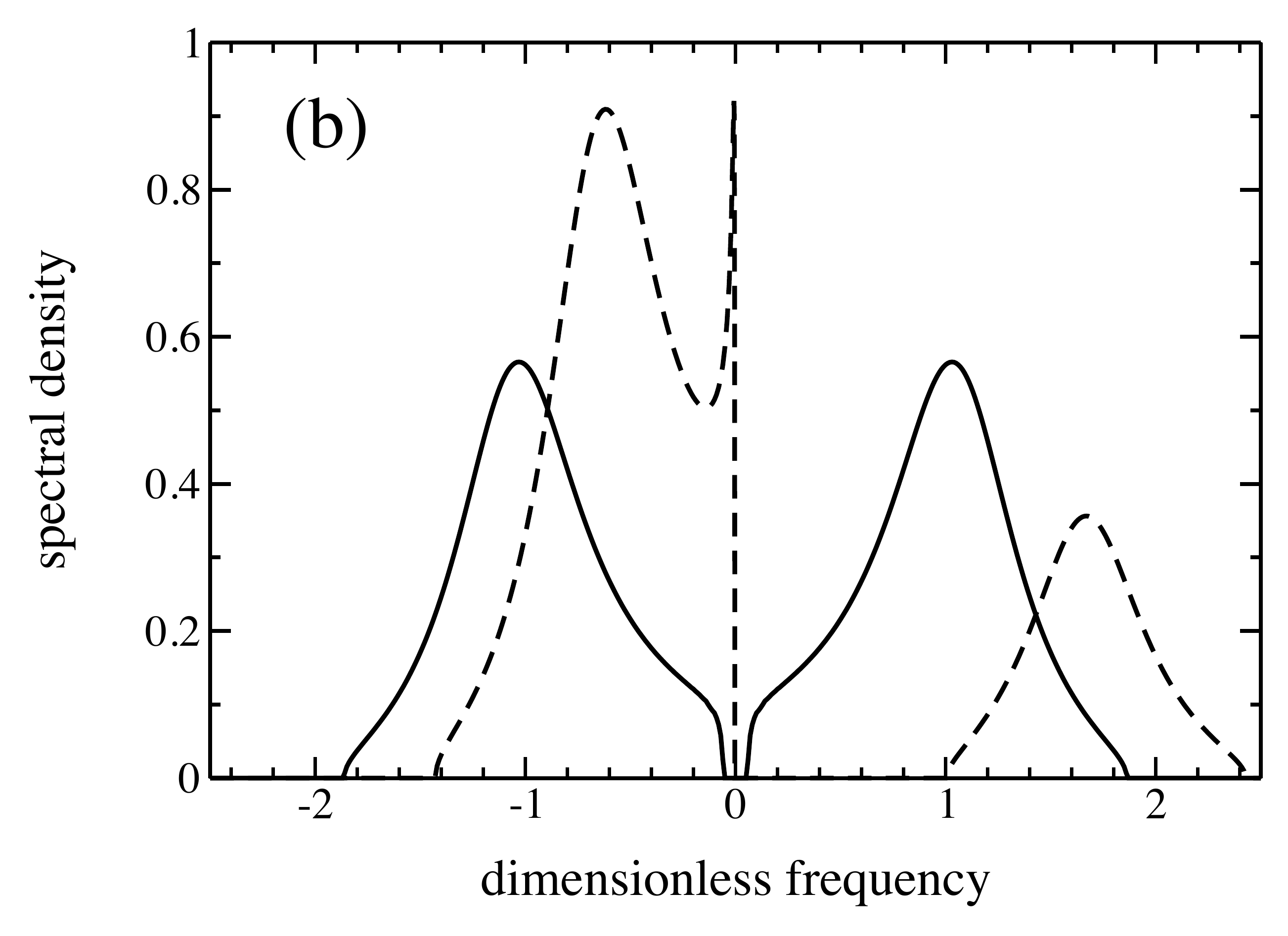}
\vspace{-0cm}
\end{center}
\caption{The dimensionless spectral densities
(a)~of the asymmetric electronic excitation,
$-(V_K/\pi)\Im\Pi_{2,\pm{1}}(\Omega)$, and
(b)~of the $\Gamma$~point phonon,
$(-V_\Gamma/\pi)\Im\tilde{D}_\Gamma(\vec{q}=0,\omega)$,
versus the dimensionless frequency
(a)~$z=(\Omega-\omega_K+E_{-9\pm{1}})/V_K$, and
(b)~$z=(\omega-\omega_\Gamma)/V_\Gamma$,
as obtained from the $GW$ approximation.
On both panels, the solid line correponds to the case of
exact resonance,
$\delta=(E_{2,1}-\omega_{K,\Gamma})/V_{K,\Gamma}=0$, the
dashed line to $\delta=1$. The spikes for $\delta=1$
do not correspond to any divergencies in the spectral
function, they are sharp but finite.
}
\label{fig:spectrInt}
\end{figure}

A qualitative description of this broadening effect can be
done by keeping only the first diagram in
Fig.~(\ref{fig:integerDiag})(b) and the first two diagrams in
Fig.~(\ref{fig:integerDiag})(d). This procedure, analogous to
the $GW$~approximation for interacting electrons (see
Ref.~\cite{Aryasetiawan1998} for a review), gives a closed
system of self-consistent equations for the dressed electron
and phonon Green's functions, $\tilde{G}$ and $\tilde{D}$.
For the coupling to the $K$~point phonons this system has
the following form:
\begin{subequations}
\begin{align}
&\tilde{D}_K^{-1}(\vec{q},\omega)
=\omega-\omega_K-4V_K^2\left[
f_0|\tilde{J}_{n0}^z(\vec{q})|^2\tilde{G}_n(\omega)\right.-\nonumber\\
&\hspace*{2cm}-\left.(1-f_0)|\tilde{J}_{0,-n}^z(\vec{q})|^2
\tilde{G}_{-n}(-\omega)\right],\\
&\tilde{G}_n^{-1}(\ep)=\ep-E_n-(1-f_0)\lambda_K{v}^2\times\nonumber\\
&\hspace*{2cm}{}\times\int\frac{d^2\vec{q}}{(2\pi)^2}\sum_\sigma
|\tilde{J}_{n0}^z(\vec{q})|^2\tilde{D}_K(\vec{q},\ep),\\
&\tilde{G}_{-n}^{-1}(-\ep)=-\ep+E_n+f_0\lambda_K{v}^2\times\nonumber\\
&\hspace*{1.8cm}{}\times\int\frac{d^2\vec{q}}{(2\pi)^2}\sum_\sigma
|\tilde{J}_{0,-n}^z(\vec{q})|^2\tilde{D}_K(\vec{q},\ep),
\end{align}
\end{subequations}
where $0<f_0<1$ is the filling of the zero Landau level.
For the resonance with $\Gamma$~point phonons, the phonons
corresponding to the two circular polarizations are dressed
differently (unless $f_0=1/2$), so the above equations should
be modified by (i)~replacing $\tilde{D}_K\to\tilde{D}_\sigma$,
$\tilde{J}^z\to\tilde{J}^\sigma$, and (ii)~introducing an
additional factor of two in the first equation,
$4V_K^2\to{8}V_K^2$, due to the valley degeneracy.
%
Restricting ourselves to the electron-hole-symmetric case 
with $f_0=1/2$, we eliminate the phonon Green's function,
and obtain a single \emph{algebraic} equation for
$\tilde{G}_n(\ep)=-\tilde{G}_{-n}(-\ep)$:
\begin{equation}\label{Gselfcons=}
\frac{1}{\tilde{G}_n(\ep)}=\ep-E_n
-\int
\frac{\pi{l}_B^2V_K^2|J_{n0}(\vec{q})|^2\,{d^2\vec{q}}/{(2\pi)^2}}%
{\ep-\omega_K-2V_K^2|J_{n0}(\vec{q})|^2\tilde{G}_n(\ep)}.
\end{equation}
The equation for the resonance with the $\Gamma$~phonon
is obtained from this one by replacements
$\omega_K\to\omega_\Gamma$, $V^2_K\to{V}^2_\Gamma$, 
$|J_{n0}(\vec{q})|^2\to|J_{n-1,0}(\vec{q})|^2$.
It is convenient to introduce dimensionless variables
$z=(\ep-\omega_K)/V_K$ and $g(z)=V_K\tilde{G}_n(\ep)$, as
well as the dimensionless detuning $\delta=(E_n-\omega_K)/V_K$.
Then the Raman spectrum for the asymmetric excitation,
$(-1/\pi)\Im\Pi_{n,\pm{1}}(\Omega)=
(-1/\pi)\Im\tilde{G}_n(\Omega-E_{n\pm{1}})
\propto(-1/\pi)\Im{g}(z)$. The latter quantity, obtained from
the numerical solution of Eq.~(\ref{Gselfcons=}) (see
Appendix~\ref{app:equation} for discussion of some properties
of this equation) is plotted in Fig.~(\ref{fig:spectrInt})
for $n=2$ and two values of detuning $\delta=0,1$.

The same discussion can be applied to the magnetophonon
resonance in the Raman scattering on the $\Gamma$-point
phonons, seen as the resonant modulation of the $G$~peak
\cite{Faugeras2009,Yan2010}. The theory developed in
Refs.~\cite{Ando2007,Goerbig2007}, when considered close
to each resonance, essentially maps on an effective two-level
problem. This is correct for transitions not involving the
zero Landau level, whose partial filling makes the picture
for the $L_{0,1}$ transition more complicated, on the same
level of difficulty as discussed above. Indeed, the Raman
spectrum in this case is proportional to the phonon
spectral function, $(-1/\pi)\Im\tilde{D}_K(\vec{q}=0,\omega)$,
given by the diagrams of Fig.~\ref{fig:integerDiag}(b). The
effect is illustrated in the same $GW$ approximation in
Fig.~\ref{fig:spectrInt}(b), where we plot the spectrum
$(-1/\pi)\Im[1/(z-g(z))]\propto
(-1/\pi)\Im\tilde{D}_\Gamma(\vec{q}=0,\omega)$, obtained from
the numerical solution of Eq.~(\ref{Gselfcons=}), modified
to describe the coupling to the $\Gamma$~phonons, using
$V_\Gamma$ as the unit of energy and introducing the shifted
frequency $z=(\omega-\omega_\Gamma)/V_\Gamma$, and for two
values of detuning $\delta=(E_1-\omega_\Gamma)/V_\Gamma=0$
and~1.

The $GW$ calculation, presented above, is not justified by
any small parameter; near the resonance, the diagrams in
Fig.~\ref{fig:integerDiag}(c,d) which are neglected, are
of the same order as those that are retained. So there is
no reason to believe that the complicated spectral shapes
shown in Fig.~\ref{fig:spectrInt} have any quantitaive
significance. (For the same reason we do not study the
symmetric excitations, $L_{-n,n}$ in the $GW$ approximation.)
Fig.~\ref{fig:spectrInt} serves only as an illustration of
the key fact: even in the simplest model used here, which
neglects many effects (such as the dispersion of phonons and
inter-Landau-level electronic excitations, their broadening
by disorder, fine structure of intra-level electronic
excitations on the zero Landau level),
the electron-phonon interaction involving the partially
filled Landau level broadens the discrete peaks into
non-trivial spectral shapes. The details of these shapes
are likely to be masked by the above-mentioned effects
anyway.

Still, we need a practical recipe for a quantitative analysis
of the experimental data. In Sec.~\ref{sec:experiment} we
have seen that the experimentally measured magnetic field
dependence of the positions of the Raman peaks' maxima is
reasonably well described by an effective two- or three-level
model, Eqs.~(\ref{Heff2=}), (\ref{Heff3=}).
To fix the effective parameters $C_1',C_1,C_2$, we match the
level shifts obtained from the effective models
(\ref{Heff3=}), (\ref{Heff2=}) away from the resonance,
$|E_n-\omega_{K,\Gamma}|\gg{V}_{K,\Gamma}$, to the expressions
for the energy shifts obtained perturbatively from the original
electron-phonon Hamiltonian of Sec.~\ref{ssec:model}.

First, let us find the correction to the energy $2E_n$
of the initially excited electron-hole pair,
\begin{equation}\label{eh=}
|eh\rangle=
\hat{c}^\dagger_{n,p,K}\hat{c}_{-n,p,K}|0\rangle,
\end{equation}
where $|0\rangle$ is one of the many degenerate ground
states with partially filled zero level. It is specified
by occupations (0~or~1) of states with different
momenta, the fraction of filled states being~$f_0$.
Strictly speaking, the optically excited state is given
by the symmetric linear combination of states~(\ref{eh=})
with all momenta~$p$ and in both valleys. However, the
energy shift depends neither on~$p$, nor on the valley,
so we do not write the whole linear combination for the
sake of compactness.
The second-order
perturbation theory in electron-phonon coupling gives
\begin{equation}\label{deltaEeh=}
\delta{E}_{eh}=
\lambda_Kv^2\int\frac{d^2\vec{q}}{(2\pi)^2}\,
\frac{|\tilde{J}^z_{n0}(\vec{q})|^2}{E_n-\omega_K}
=\frac{V_K^2}{E_n-\omega_K},
\end{equation}
for the coupling to the $K$~point phonons.
For the $\Gamma$~point phonons one should replace
$V_K\to{V}_\Gamma$, $\omega_K\to\omega_\Gamma$.
At the same time, the shift from the effective model
(\ref{Heff3=}) is $(C_1v/l_B)^2/(E_n-\omega_{K,\Gamma})$,
which fixes $C_1=\sqrt{\lambda_{K,\Gamma}/(4\pi)}$ for
symmetric transitions. 
The shift~(\ref{deltaEeh=}) for the symmetric excitation
comes from the coupling to states with one phonon, which
can be emitted either by the electron or by the hole. For
asymmetric transitions only one of these processes is
allowed. Thus, the energy shift is given by the same
Eq.~(\ref{deltaEeh=}), multiplied by $1-f_0$ if the
initial excitation involved level~$n$, and by $f_0$ if it
was $-n$, so the coupling matrix element of the effective
two-level model~(\ref{Heff2=}) is
$C_1'=\sqrt{(1-f_0)\lambda_{K,\Gamma}/(4\pi)}$ or
$C_1'=\sqrt{f_0\lambda_{K,\Gamma}/(4\pi)}$.

To find the effective matrix element $V_2$, we have to find
the energy shift in the one-phonon sector.
The states in the one-phonon sector are characterized by the
phonon wave vector~$\vec{q}$ and have the form
\begin{subequations}\begin{align}\label{1phK=}
&|e,\vec{q}\rangle=
\hat{c}^\dagger_{n,p,K}\hat{b}^\dagger_{\vec{q},K}|0_e\rangle,\\
&|e,\vec{q},\sigma\rangle=
\hat{c}^\dagger_{n,p,K}\hat{b}^\dagger_{\vec{q},\sigma}
|0_e\rangle,\label{1phG=}
\end{align}\end{subequations}
for the $K$ and $\Gamma$~point phonons, respectively.
The states $|0_e\rangle$ represents a ground state
with partially filled zero level, but it may be different
from the state $|0\rangle$ before the initial excitation.
In addition to states (\ref{1phK=}), (\ref{1phG=}), which
correspond to phonon emission accompanied by the hole
annihilation, there are also states corresponding to
electron annihilation obtained by replacement
$\hat{c}^\dagger_{n,p,K}\to\hat{c}_{-n,p,K}$, and of
the label ``$e$'' by~``$h$''.

The second-order correction to the the energies of states
(\ref{1phK=}), (\ref{1phG=}) is given by
\begin{subequations}\begin{align}
&\delta{E}_{e,\vec{q}}=\frac{V_K^2}{E_n-\omega_K}
\left[(1-f_0)-|{J}_{n0}(\vec{q})|^2\right],\label{deltaEz=}\\
&\delta{E}_{e,\vec{q},+}=
\frac{V_\Gamma^2}{E_n-\omega_\Gamma}
\left[(1-f_0)-f_0|J_{n-1,0}(\vec{q})|^2\right],\\
&\delta{E}_{e,\vec{q},-}=
\frac{V_\Gamma^2}{E_n-\omega_\Gamma}
\left[(1-f_0)-(1-f_0)|J_{n-1,0}(\vec{q})|^2\right].
\label{deltaEm=}
\end{align}\end{subequations}
The first term in the square brackets comes from coupling
to the two-phonon sector by annihilation of the remaining
electron, which requires an empty state on the zero level.
This gives the factor $1-f_0$, which should be replaced by
$f_0$ for the corrections $\delta{E}_{h,\vec{q}}$,
$\delta{E}_{h,\vec{q},\sigma}$. The second term is due to
coupling to the zero-phonon sector, when the phonon can be
absorbed either by producing a hole on the level $-n$ or a
second electron on the level~$n$. For the $K$~point phonon
both these process are allowed, so the occupation $f_0$
cancels out, while for the $\Gamma$~point phonons either
one or the other process is allowed for each of the two
circular polarizations $\sigma=\pm$. Obviously, for the
corrections $\delta{E}_{h,\vec{q}}$,
$\delta{E}_{h,\vec{q},\sigma}$ the second factor is the
same as in Eqs.~(\ref{deltaEz=})--(\ref{deltaEm=}).

The energy shifts are different for states with different
phonon wave vectors~$\vec{q}$, which is another
manifestation of the effect that has already been discussed:
the energy spectrum of the coupled system does not reduce
to a few discrete levels, but is continuous. 
Thus, we have to decide which energy in this continuum
should be associated with the discrete state of the effective
three-level model~(\ref{Heff3=}). Clearly, some values of
$\vec{q}$ are more representative than others, as they have
larger weight in the Raman spectrum.
This weight is determined by the overlap of the exact
eigenstates with the initial $|eh\rangle$ state.
Away from the resonance, it can be determined in the first
order of the perturbation theory in the electron-phonon
coupling.
For the coupling to the $K$~point phonons, the weight of the
states $|e,\vec{q}\rangle$, $|h,\vec{q}\rangle$ is given by
\begin{subequations}\begin{align}
&|\langle{eh}|e,\vec{q}\rangle|^2=
f_0\,\frac{2\pi{l}_B^2}{L_xL_y}\,
\frac{V_K^2|J_{n0}(\vec{q})|^2}{(E_n-\omega_K)^2},\\
&|\langle{eh}|h,\vec{q}\rangle|^2=
(1-f_0)\,\frac{2\pi{l}_B^2}{L_xL_y}\,
\frac{V_K^2|J_{n0}(\vec{q})|^2}{(E_n-\omega_K)^2},
\end{align}
while for the $\Gamma$~point phonons we have
\begin{align}
&|\langle{eh}|e,\vec{q},-\rangle|^2=
f_0\,\frac{2\pi{l}_B^2}{L_xL_y}\,
\frac{V_\Gamma^2|J_{n-1,0}(\vec{q})|^2}{(E_n-\omega_\Gamma)^2},\\
&|\langle{eh}|h,\vec{q},+\rangle|^2=
(1-f_0)\,\frac{2\pi{l}_B^2}{L_xL_y}\,
\frac{V_\Gamma^2|J_{n-1,0}(\vec{q})|^2}{(E_n-\omega_\Gamma)^2},\\
&|\langle{eh}|h,\vec{q},-\rangle|^2=
|\langle{eh}|e,\vec{q},+\rangle|^2=0.
\end{align}\end{subequations}
We choose the ``center of mass'' as the representative energy of
the continuum:
\begin{subequations}\begin{align}
\overline{\delta{E}}={}&{}
\frac{\sum_\vec{q}\delta{E}_{e,\vec{q}}|\langle{eh}|e,\vec{q}\rangle|^2
+\sum_\vec{q}\delta{E}_{h,\vec{q}}|\langle{eh}|h,\vec{q}\rangle|^2}%
{\sum_\vec{q}|\langle{eh}|e,\vec{q}\rangle|^2
+\sum_\vec{q}|\langle{eh}|h,\vec{q}\rangle|^2}=\nonumber\\
={}&{}\frac{V_K^2}{E_n-\omega_K}
\left[2f_0(1-f_0)-\frac{(2n)!}{2^{2n+1}(n!)^2}\right],\\
\overline{\delta{E}_+}={}&{}
\frac{f_0V_\Gamma^2}{E_n-\omega_\Gamma}
\left[1-\frac{(2n-2)!}{2^{2n-1}((n-1)!)^2}\right],\\
\overline{\delta{E}_-}={}&{}
\frac{(1-f_0)V_\Gamma^2}{E_n-\omega_\Gamma}
\left[1-\frac{(2n-2)!}{2^{2n-1}((n-1)!)^2}\right].
\end{align}\end{subequations}
These shifts should be matched with that obtained from the
effective model~(\ref{Heff3=}),
$(C_2^2-C_1^2)(v/l_B)^2/(E_n-\omega_{K,\Gamma})$.
The resulting values of $C_2$ for the first three transitions at 
half-filling ($f_0=1/2$) are given in Table~\ref{tab:couplTh}.
They are compared to the experimental values in 
Table~\ref{tab:couplExpTh} (end of Sec.~\ref{sec:experiment}).

\begin{table}
\begin{tabular}{|c|c|c|c|}
\hline transition & resonance & $C_1$ & $C_2$ \\
\hline $L_{-1,1}$ & $E_1=\omega_K$ & 
$0.282\,\sqrt{\lambda_K}$  & $0.315\,\sqrt{\lambda_K}$ \\
\hline $L_{-1,1}$ & $E_1=\omega_\Gamma$ & 
$0.282\,\sqrt{\lambda_\Gamma}$  & $0.315\,\sqrt{\lambda_\Gamma}$ \\
\hline $L_{-2,2}$ & $E_2=\omega_K$ & 
$0.282\,\sqrt{\lambda_K}$  & $0.323\,\sqrt{\lambda_K}$ \\
\hline $L_{-2,2}$ & $E_2=\omega_\Gamma$ & 
$0.282\,\sqrt{\lambda_\Gamma}$  & $0.331\,\sqrt{\lambda_\Gamma}$ \\
\hline $L_{-3,3}$ & $E_3=\omega_K$ & 
$0.282\,\sqrt{\lambda_K}$  & $0.327\,\sqrt{\lambda_K}$ \\
\hline $L_{-3,3}$ & $E_3=\omega_\Gamma$ & 
$0.282\,\sqrt{\lambda_\Gamma}$  & $0.335\,\sqrt{\lambda_\Gamma}$ \\
\hline
\end{tabular}
\caption{The dimensionless couplings $C_1,C_2$ in the effective 
three-level model~(\ref{Heff3=}) for the three lowest symmetric
electronic transitions $L_{-1,1}$, $L_{-2,2}$, $L_{-3,3}$ and
$K$ and $\Gamma$ point phonons at half-filling of the zero Landau
level ($f_0=1/2$). For the effective two-level model~(\ref{Heff2=}), 
we then have $C_1'=C_1/\sqrt{2}$.
}
\label{tab:couplTh}
\end{table}

\section{Conclusions}

To conclude, we have presented a complete theoretical and experimental picture of magneto-phonon resonances in graphene, as observed in the magneto-Raman scattering spectra. By considering the case of an electronic inter-Landau-level excitation decaying into a phonon and another electronic excitation, we have extended the commonly accepted picture of magneto-phonon resonance in graphene (i)~to both $\Gamma$ and $K$~point phonons and (ii)~to multiphonon processes. Such multi-excitation final states have numerous possibilities to fulfill momentum conservation, including intervalley electronic excitations. We have derived a simple classification of such magneto-phonon resonances and we have described the properties of each different type of resonances. In particular, resonant splitting of three crossing excitation branches is observed experimentally for the first time. Such multi-phonon processes are particularly important for the understanding of hot carrier relaxation in graphene. 

We analyze the experimental data phenomenologically in terms of effective two- or three-level models, and then give a detailed discussion of validity of such effective description starting from a microscopic theory in the resonant approximation. We also highlight the richness of the physics associated with the partially filled zero Landau level and its low-energy electronic excitations. When this level is involved in a resonance, the possibility to excite an arbitrary number of intra-level excitations with zero energy (i)~leads to a genuine many-body problem which is quite hard to analyze theoretically, and (ii)~results in complicated spectral shapes of the split peaks.

\begin{acknowledgments}
We thank M. O. Goerbig and J.-N. Fuchs for fruitful discussions.
This work was supported by
\end{acknowledgments}

\appendix

\section{Magnetopolaron wave functions}
\label{app:wavefunction}

Let us first consider the case of coupling to $K,K'$~point
phonons, which results in Eq.~(\ref{Pinovertex=}).
Let us look for the eigenstates in the form
\begin{align}\label{PsiK=}
&|\Psi\rangle=\Psi^{eh}|eh\rangle
+\sum_{\vec{q}}\Psi^e_{\vec{q}}|e,\vec{q}\rangle
+\sum_{\vec{q}}\Psi^h_{\vec{q}}|h,\vec{q}\rangle
+{}\nonumber\\ &\qquad\quad{}
+\sum_{\vec{q}_1,\vec{q}_2}\Psi_{\vec{q}_1,\vec{q}_2}
|\vec{q}_1,\vec{q}_2\rangle,
\end{align}
where the basis states are defined as
\begin{subequations}\begin{align}
|eh\rangle={}&{}\sqrt{\frac{2\pi{l}_B^2}{L_xL_y}}\sum_p
\hat{c}^\dagger_{n,p,K}\hat{c}_{-n,p,K}|0\rangle,\\
|e,\vec{q}\rangle={}&{}\sqrt{\frac{2\pi{l}_B^2}{L_xL_y}}
\sum_p e^{ipq_yl_B^2}
\hat{c}^\dagger_{n,p-q_x/2,K}\hat{c}_{-n',p+q_x/2,K'}{}\nonumber\\ 
{}&{}\times\hat{b}^\dagger_{\vec{q},K}|0\rangle,\\
|h,\vec{q}\rangle={}&{}\sqrt{\frac{2\pi{l}_B^2}{L_xL_y}}
\sum_p e^{ipq_yl_B^2}
\hat{c}^\dagger_{n',p-q_x/2,K'}\hat{c}_{-n,p+q_x/2,K}{}\nonumber\\ 
{}&{}\times\hat{b}^\dagger_{\vec{q},K'}|0\rangle,\\
|\vec{q}_1,\vec{q}_2\rangle={}&{}
\sqrt{\frac{2\pi{l}_B^2}{L_xL_y}}\sum_p e^{ipQ_yl_B^2}
\hat{c}^\dagger_{n',p-Q_x/2,K'}\hat{c}_{-n',p+Q_x/2,K'}\nonumber\\
{}&{}\times\hat{b}^\dagger_{\vec{q}_1,K}\hat{b}^\dagger_{\vec{q}_2,K'}
|0\rangle,\quad \vec{Q}\equiv\vec{q}_1+\vec{q}_2.
\end{align}\end{subequations}
Here $|0\rangle$ is the ground state.
Note that, generally speaking,
$\Psi_{\vec{q}_1,\vec{q}_2}\neq\Psi_{\vec{q}_2,\vec{q}_1}$
as the phonons belong to different valleys and thus are
effectively distinguishable in the two-phonon state
$|\vec{q}_1,\vec{q}_2\rangle$.

Eigenstates of the form (\ref{PsiK=}) belong to the sector where
the initial electron-hole pair $|eh\rangle$ is in the $K$ valley.
There are also eigenstates in the complementary sector, obtained
by exchanging $K\leftrightarrow{K}'$ in the above expressions,
with exactly the same energies, as the Hamiltonian has the overall
valley symmetry. The optical excitation corresponds to the
symmetric combination of states from the two sectors.


The Schr\"odinger equation $\hat{H}|\Psi\rangle=\Omega|\Psi\rangle$
results in the following system of linear equations for the
coefficients in Eq.~(\ref{PsiK=}):
\begin{subequations}\begin{align}
&\sqrt{\frac{L_xL_y}{\lambda_K{v}^2}}\,(\Omega-2E_n)\Psi^{eh}=\nonumber\\
{}&{}=-\sum_{\vec{q}}\tilde{J}^z_{-n',-n}(\vec{q})\,\Psi^e_\vec{q}
+\sum_{\vec{q}}\tilde{J}^z_{nn'}(\vec{q})\,\Psi^h_\vec{q},
\label{SchrK1=}\\
&\sqrt{\frac{L_xL_y}{\lambda_K{v}^2}}\,
(\Omega-E_n-E_{n'}-\omega_K)\,\Psi^e_\vec{q}=\nonumber\\
{}&{}=-[\tilde{J}^z_{-n',-n}(\vec{q})]^*\Psi^{eh}+
\sum_{\vec{q}'}\tilde{J}^z_{nn'}(\vec{q}')\,
e^{-i[\vec{q}'\times\vec{q}]_zl_B^2/2}\Psi_{\vec{q},\vec{q}'},\\
&\sqrt{\frac{L_xL_y}{\lambda_K{v}^2}}\,
(\Omega-E_n-E_{n'}-\omega_K)\,\Psi^h_\vec{q}=\nonumber\\
{}&{}=[\tilde{J}^z_{nn'}(\vec{q})]^*\Psi^{eh}
-\sum_{\vec{q}'}\tilde{J}^z_{-n',-n}(\vec{q}')\,
e^{i[\vec{q}'\times\vec{q}]_zl_B^2/2}\Psi_{\vec{q}',\vec{q}},\\
&\sqrt{\frac{L_xL_y}{\lambda_K{v}^2}}\,
(\Omega-2E_{n'}-2\omega_K)\Psi_{\vec{q}_1,\vec{q}_2}=\nonumber\\
{}&{}={}
[\tilde{J}^z_{nn'}(\vec{q}_2)]^*e^{i[\vec{q}_2\times\vec{q}_1]_zl_B^2/2}
\Psi^e_{\vec{q}_1}-{}\nonumber\\
&\quad{}-[\tilde{J}^z_{-n',-n}(\vec{q}_1)]^*
e^{i[\vec{q}_2\times\vec{q}_1]_zl_B^2/2}\Psi^h_{\vec{q}_2}.
\end{align}\end{subequations}
Let us look for the solution in the separable form
\begin{subequations}\begin{align}
&\Psi^e_\vec{q}=-\sqrt{\frac{\lambda_K{v}^2}{L_xL_y}}\,
\frac{[\tilde{J}^z_{-n',-n}(\vec{q})]^*}{V_K}\,\Psi^e,\\
&\Psi^h_\vec{q}=\sqrt{\frac{\lambda_K{v}^2}{L_xL_y}}\,
\frac{[\tilde{J}^z_{nn'}(\vec{q})]^*}{V_K}\,\Psi^h,\\
&\Psi_{\vec{q}_1,\vec{q}_2}=-\frac{\lambda_K{v}^2}{L_xL_y}\,
\frac{[\tilde{J}^z_{-n',-n}(\vec{q}_1)]^*
[\tilde{J}^z_{nn'}(\vec{q}_2)]^*}{V_K^2}\times{}\nonumber\\
&\qquad\qquad{}\times{}
e^{i[\vec{q}_2\times\vec{q}_1]_zl_B^2/2}\Psi^{(2)}.
\label{SchrK4=}
\end{align}\end{subequations}
Then the normalization condition is
\begin{equation}
\langle\Psi|\Psi\rangle=
|\Psi^{eh}|^2+|\Psi^e|^2+|\Psi^h|^2+|\Psi^{(2)}|^2=1,
\end{equation}
and the system of linear equations can be written as
\begin{equation}
\left(\Omega-H_\mathrm{eff}^{4\times{4}}\right)
\left(\begin{array}{c} \Psi^{eh} \\ \Psi^e \\ \Psi^h \\ \Psi^{(2)}
\end{array}\right)=0,
\end{equation}
where the effective Hamiltonian matrix is given by
\begin{equation}
H_\mathrm{eff}^{4\times{4}}=
\left(\begin{array}{cccc} 2E_n & V_K & V_K & 0 \\ 
V_K & E_n+E_n' & 0 & V_K \\ V_K & 0 & E_n+E_n' & V_K \\
0 & V_K & V_K & 2E_n' \end{array}\right).
\end{equation}
The upper left matrix element of 
$(\Omega-H_\mathrm{eff}^{4\times{4}})^{-1}$ coincides
with Eq.~(\ref{Pinovertex=}), which proves their equivalence.

For the coupling to $\Gamma$~point phonons, we can look for the
eigenstate in the form 
\begin{align}
|\Psi\rangle={}&{}\Psi^{eh}|eh\rangle
+\sum_{\vec{q},\sigma}\Psi^e_{\vec{q},\sigma}
|e,\vec{q},\sigma\rangle
+\sum_{\vec{q},\sigma}\Psi^h_{\vec{q},\sigma}
|h,\vec{q},\sigma\rangle 
+{}\nonumber\\ &\qquad\quad{}
+\frac{1}{2}\sum_{\vec{q}_1,\sigma_1,\vec{q}_2,\sigma_2}
\Psi_{\vec{q}_1,\sigma_1,\vec{q}_2,\sigma_2}
|\vec{q}_1,\sigma_1,\vec{q}_2,\sigma_2\rangle,\label{PsiG=}
\end{align}
where $\Psi_{\vec{q}_1,\sigma_1,\vec{q}_2,\sigma_2}
=\Psi_{\vec{q}_2,\sigma_2,\vec{q}_1,\sigma_1}$.
The basis states are defined as
\begin{subequations}\begin{align}
&|eh\rangle=\sqrt{\frac{2\pi{l}_B^2}{L_xL_y}}\sum_p
\hat{c}^\dagger_{n,p,K}\hat{c}_{-n,p,K}|0\rangle,\\
&|e,\vec{q},\sigma\rangle=\sqrt{\frac{2\pi{l}_B^2}{L_xL_y}}
\sum_p e^{ipq_yl_B^2}
\times\nonumber\\ &\qquad\qquad{}\times
\hat{c}^\dagger_{n,p-q_x/2,K}\hat{c}_{-n',p+q_x/2,K}
\hat{b}^\dagger_{\vec{q},\sigma}|0\rangle,\\
&|h,\vec{q},\sigma\rangle=\sqrt{\frac{2\pi{l}_B^2}{L_xL_y}}
\sum_p e^{ipq_yl_B^2}
\times\nonumber\\ &\qquad\qquad{}\times
\hat{c}^\dagger_{n',p-q_x/2,K}\hat{c}_{-n,p+q_x/2,K}
\hat{b}^\dagger_{\vec{q},\sigma}|0\rangle,\\
&|\vec{q}_1,\sigma_1,\vec{q}_2,\sigma_2\rangle
=\sqrt{\frac{2\pi{l}_B^2}{L_xL_y}}
\sum_p e^{ipQ_yl_B^2}
\times\nonumber\\ &\qquad\qquad{}\times
\hat{c}^\dagger_{n',p-Q_x/2,K}\hat{c}_{-n',p+Q_x/2,K}
\hat{b}^\dagger_{\vec{q}_1,\sigma_1}
\hat{b}^\dagger_{\vec{q}_2,\sigma_2}|0\rangle.
\end{align}\end{subequations}
Again, in the last line $\vec{Q}\equiv\vec{q}_1+\vec{q}_2$,
and we focuse on the sector where the initial electron-hole
pair $|eh\rangle$ is in the $K$ valley. Note that
$|\vec{q}_1,\sigma_1,\vec{q}_2,\sigma_2\rangle$ and
$|\vec{q}_2,\sigma_2,\vec{q}_1,\sigma_1\rangle$ represent the
same state, so the wave function in the two-phonon sector is
assumed to be symmetric,
$\Psi_{\vec{q}_1,\sigma_1,\vec{q}_2,\sigma_2}=
\Psi_{\vec{q}_2,\sigma_2,\vec{q}_1,\sigma_1}$,
and the factor 1/2 is necessary in Eq.~(\ref{PsiG=}).

The Schr\"odinger equation $\hat{H}|\Psi\rangle=\Omega|\Psi\rangle$
results in the following system of linear equations for the
coefficients in Eq.~(\ref{PsiG=}):
\begin{subequations}\begin{align}
&\sqrt{\frac{L_xL_y}{\lambda_\Gamma{v}^2}}\,(\Omega-2E_n)\Psi^{eh}
=\nonumber\\{}&{}=
-\sum_{\vec{q},\sigma}\tilde{J}^\sigma_{-n',-n}(\vec{q})\,
\Psi^e_{\vec{q},\sigma}
+\sum_{\vec{q},\sigma}\tilde{J}^\sigma_{nn'}(\vec{q})\,
\Psi^h_{\vec{q},\sigma},\label{SchrGeh=}\\
&\sqrt{\frac{L_xL_y}{\lambda_\Gamma{v}^2}}\,
(\Omega-E_n-E_{n'}-\omega_\Gamma)\Psi^e_{\vec{q},\sigma}
=\nonumber\\{}&{}=
-[\tilde{J}^\sigma_{-n',-n}(\vec{q})]^*\Psi^{eh}+{}\nonumber\\
&\quad{}+
\sum_{\vec{q}',\sigma'}\tilde{J}^{\sigma'}_{nn'}(\vec{q}')\,
e^{-i[\vec{q}'\times\vec{q}]_zl_B^2/2}
\Psi_{\vec{q}',\sigma',\vec{q},\sigma},\label{SchrGe=}\\
&\sqrt{\frac{L_xL_y}{\lambda_\Gamma{v}^2}}\,
(\Omega-E_n-E_{n'}-\omega_\Gamma)\Psi^h_{\vec{q},\sigma}
=\nonumber\\{}&{}=
[\tilde{J}^\sigma_{nn'}(\vec{q})]^*\Psi^{eh}-{}\nonumber\\
&\quad{}-\sum_{\vec{q}',\sigma'}
\tilde{J}^{\sigma'}_{-n',-n}(\vec{q}')\,
e^{i[\vec{q}'\times\vec{q}]_zl_B^2/2}
\Psi_{\vec{q}',\sigma',\vec{q},\sigma},\label{SchrGh=}\\
&\sqrt{\frac{L_xL_y}{\lambda_\Gamma{v}^2}}\,
(\Omega-2E_{n'}-2\omega_\Gamma)
\Psi_{\vec{q}_1,\sigma_1,\vec{q}_2,\sigma_2}
=\nonumber\\{}&{}=
[\tilde{J}^{\sigma_1}_{nn'}(\vec{q}_1)]^*\,
e^{i[\vec{q}_1\times\vec{q}_2]l_B^2/2}\Psi^e_{\vec{q}_2,\sigma_2}
+\nonumber\\
&\quad{}+[\tilde{J}^{\sigma_2}_{nn'}(\vec{q}_2)]^*\,
e^{i[\vec{q}_2\times\vec{q}_1]l_B^2/2}\Psi^e_{\vec{q}_1,\sigma_1}
-\nonumber\\
&\quad{}-[\tilde{J}^{\sigma_1}_{-n',-n}(\vec{q}_1)]^*\,
e^{i[\vec{q}_2\times\vec{q}_1]l_B^2/2}\Psi^h_{\vec{q}_2,\sigma_2}
-\nonumber\\
&\quad{}-[\tilde{J}^{\sigma_2}_{-n',-n}(\vec{q}_2)]^*\,
e^{i[\vec{q}_1\times\vec{q}_2]l_B^2/2}\Psi^h_{\vec{q}_1,\sigma_1}.
\label{SchrG2=}
\end{align}\end{subequations}
This system differs from Eqs. (\ref{SchrK1=})--(\ref{SchrK4=})
by the presence of two extra terms in the last equation, due to
the symmetry of the wave function in the two-phonon sector.
These terms correspond to the exchange of phonons between the
electron and the hole, and because of them the whole system
becomes non-separable. 

\section{Some properties of Eq.~(\ref{Gselfcons=})}
\label{app:equation}

In the dimensionless variables $z=(\ep-\omega_K)/V_K$,
$g=V_K\tilde{G}_n$, $\delta=(E_n-\omega_K)/V_K$,
Eq.~(\ref{Gselfcons=}) acquires the following form:
\begin{equation}\label{1g=}
\frac{1}{g}=z-\delta-\frac{1}{2g}\int\limits_0^\infty
\frac{\xi^ne^{-\xi}/n!}{z/g-2\xi^ne^{-\xi}/n!}\,d\xi.
\end{equation}
The function
\begin{equation}\label{Fnu=}
\mathcal{F}_n(u)\equiv\int\limits_0^\infty
\frac{\xi^ne^{-\xi}/n!}{u-2\xi^ne^{-\xi}/n!+i0^+}\,d\xi
\end{equation}
is real unless $0<u<{u}_n^{max}\equiv{2}n^ne^{-n}/n!$.
It behaves as $1/u$ at $|u|\to\infty$. At $u\to-0^+$ it
diverges as $-\xi_+(|u|)$, the larger of the solutions
of the equation $|u|=2\xi^ne^{-\xi}/n!$.
At $u\to{u}_n^{max}$, it has a square-root singularity,
provided that $n>0$. For $n=0$,it can be evaluated
explicitly: $\mathcal{F}_0(u)=(1/2)\ln[u/(u-2)]$.

Treating $u=z/g$ as a new unknown variable, we rewrite
Eq.~(\ref{1g=}) as
\begin{equation}\label{1Fnu=}
\frac{1}{\mathcal{F}_n(u)}=\frac{u/2}{z(z-\delta)-u},
\end{equation}
so that at any fixed $n$, it contains only one parameter
$w=z(z-\delta)$. Numerical solution of Eq.~(\ref{1Fnu=})
is facilitated by noting the following properties.
\begin{itemize}
\item
At large positive $w$, corresponding to large $|z|$ or
large negative detuning~$\delta$, Eq.~(\ref{1Fnu=}) has
a real solution $u\approx{z}(z-\delta)$, corresponding to
the unperturbed Green's function $g(z)=1/(z-\delta)$.
Another real solution is at $u\approx{u}_n^{max}$.
\item
As $w$ is decreased below some critical value $w_n^{max}$,
the two real solutions merge, and two complex solutions
appear [the sign of $\Im{u}$ should be chosen to give
the correct analytical properties of $g(z)$].
\item
For $w$ large and negative (i.~e., moderate~$|z|$ and
large positive $\delta$), there is also a real solution
$u\approx{z}(z-\delta)<0$, as well as another one close
to $u=0$. The two solutions merge when $w=w_n^{min}$.
It is easy to see that for $w=0$ real solution exist,
so $w_n^{min}>0$.
\item
In the region $w_n^{min}<z(z-\delta)<w_n^{max}$ the
solution is complex, which means that there are two
disconnected intervals of $z$ where the spectral
density is non-zero.
\end{itemize}
Specifically, for $n=2$ we have
$w_2^{min}=0.0296\ldots$, $w_2^{max}=1.8265\ldots$,
the corresponding values of $u$ are 
$u=-0.070\ldots$ and $u=0.901\ldots$.
For $n=0$, $w_0^{min}$ and $w_0^{max}$ are the solutions
of the equation $\ln{2}w=2w-5$, 
$w_0^{min}=0.00339\ldots$, $w_0^{max}=3.4684\ldots$;
the corresponding values of $u$ are $u=4w_0/(2w_0-1)$
which gives $-0.0137\ldots$ and $2.337\ldots$. 
The numerical smallness of $w_n^{min}$ gives rise to
sharp but finite features in the spectral density.
For example, the shape of the spike in
Fig.~\ref{fig:spectrInt}(b) is given by
\begin{equation}
-\frac{1}\pi\Im\frac{1}{z-g(z)}\propto
\frac{\sqrt{-w_0^{min}-z\delta}}{|z|}.
\end{equation}
The typical width of the spike is
$\sim{w}_0^{min}/\delta$, which appears very narrow
on the scale of the figure.


\begin{thebibliography}{37}%
\makeatletter
\providecommand \@ifxundefined [1]{%
 \@ifx{#1\undefined}
}%
\providecommand \@ifnum [1]{%
 \ifnum #1\expandafter \@firstoftwo
 \else \expandafter \@secondoftwo
 \fi
}%
\providecommand \@ifx [1]{%
 \ifx #1\expandafter \@firstoftwo
 \else \expandafter \@secondoftwo
 \fi
}%
\providecommand \natexlab [1]{#1}%
\providecommand \enquote  [1]{``#1''}%
\providecommand \bibnamefont  [1]{#1}%
\providecommand \bibfnamefont [1]{#1}%
\providecommand \citenamefont [1]{#1}%
\providecommand \href@noop [0]{\@secondoftwo}%
\providecommand \href [0]{\begingroup \@sanitize@url \@href}%
\providecommand \@href[1]{\@@startlink{#1}\@@href}%
\providecommand \@@href[1]{\endgroup#1\@@endlink}%
\providecommand \@sanitize@url [0]{\catcode `\\12\catcode `\$12\catcode
  `\&12\catcode `\#12\catcode `\^12\catcode `\_12\catcode `\%12\relax}%
\providecommand \@@startlink[1]{}%
\providecommand \@@endlink[0]{}%
\providecommand \url  [0]{\begingroup\@sanitize@url \@url }%
\providecommand \@url [1]{\endgroup\@href {#1}{\urlprefix }}%
\providecommand \urlprefix  [0]{URL }%
\providecommand \Eprint [0]{\href }%
\providecommand \doibase [0]{http://dx.doi.org/}%
\providecommand \selectlanguage [0]{\@gobble}%
\providecommand \bibinfo  [0]{\@secondoftwo}%
\providecommand \bibfield  [0]{\@secondoftwo}%
\providecommand \translation [1]{[#1]}%
\providecommand \BibitemOpen [0]{}%
\providecommand \bibitemStop [0]{}%
\providecommand \bibitemNoStop [0]{.\EOS\space}%
\providecommand \EOS [0]{\spacefactor3000\relax}%
\providecommand \BibitemShut  [1]{\csname bibitem#1\endcsname}%
\let\auto@bib@innerbib\@empty
\bibitem [{\citenamefont {Yoshioka}(2010)}]{Yoshioka2010}%
  \BibitemOpen
  \bibfield  {author} {\bibinfo {author} {\bibfnamefont {D.}~\bibnamefont
  {Yoshioka}},\ }\href@noop {} {\emph {\bibinfo {title} {The Quantum Hall
  Effect}}}\ (\bibinfo  {publisher} {Springer},\ \bibinfo {year}
  {2010})\BibitemShut {NoStop}%
\bibitem [{\citenamefont {Gurevich}\ and\ \citenamefont
  {Firsov}(1961)}]{Gurevich1961}%
  \BibitemOpen
  \bibfield  {author} {\bibinfo {author} {\bibfnamefont {V.~L.}\ \bibnamefont
  {Gurevich}}\ and\ \bibinfo {author} {\bibfnamefont {Y.~A.}\ \bibnamefont
  {Firsov}},\ }\href@noop {} {\bibfield  {journal} {\bibinfo  {journal} {Sov.
  Phys. JETP}\ }\textbf {\bibinfo {volume} {13}},\ \bibinfo {pages} {137}
  (\bibinfo {year} {1961})}\BibitemShut {NoStop}%
\bibitem [{\citenamefont {Puri}\ and\ \citenamefont
  {Geballe}(1963)}]{Puri1963}%
  \BibitemOpen
  \bibfield  {author} {\bibinfo {author} {\bibfnamefont {S.~M.}\ \bibnamefont
  {Puri}}\ and\ \bibinfo {author} {\bibfnamefont {T.~H.}\ \bibnamefont
  {Geballe}},\ }\href@noop {} {\bibfield  {journal} {\bibinfo  {journal} {Bull.
  Am. Phys. Soc.}\ }\textbf {\bibinfo {volume} {8}},\ \bibinfo {pages} {309}
  (\bibinfo {year} {1963})}\BibitemShut {NoStop}%
\bibitem [{\citenamefont {Shalyt}\ \emph {et~al.}(1964)\citenamefont {Shalyt},
  \citenamefont {Parfen'ev},\ and\ \citenamefont {Muzhdaba}}]{Shalyt1964}%
  \BibitemOpen
  \bibfield  {author} {\bibinfo {author} {\bibfnamefont {S.~S.}\ \bibnamefont
  {Shalyt}}, \bibinfo {author} {\bibfnamefont {R.~V.}\ \bibnamefont
  {Parfen'ev}}, \ and\ \bibinfo {author} {\bibfnamefont {V.~M.}\ \bibnamefont
  {Muzhdaba}},\ }\href@noop {} {\bibfield  {journal} {\bibinfo  {journal} {Sov.
  Phys. Solid State}\ }\textbf {\bibinfo {volume} {6}},\ \bibinfo {pages} {508}
  (\bibinfo {year} {1964})}\BibitemShut {NoStop}%
\bibitem [{\citenamefont {Johnson}\ and\ \citenamefont
  {Larsen}(1966)}]{Johnson1966}%
  \BibitemOpen
  \bibfield  {author} {\bibinfo {author} {\bibfnamefont {E.~J.}\ \bibnamefont
  {Johnson}}\ and\ \bibinfo {author} {\bibfnamefont {D.~M.}\ \bibnamefont
  {Larsen}},\ }\href {\doibase 10.1103/PhysRevLett.16.655} {\bibfield
  {journal} {\bibinfo  {journal} {Phys. Rev. Lett.}\ }\textbf {\bibinfo
  {volume} {16}},\ \bibinfo {pages} {655} (\bibinfo {year} {1966})}\BibitemShut
  {NoStop}%
\bibitem [{\citenamefont {Korovin}\ and\ \citenamefont
  {Pavlov}(1968)}]{Korovin1968}%
  \BibitemOpen
  \bibfield  {author} {\bibinfo {author} {\bibfnamefont {L.~I.}\ \bibnamefont
  {Korovin}}\ and\ \bibinfo {author} {\bibfnamefont {S.~T.}\ \bibnamefont
  {Pavlov}},\ }\href@noop {} {\bibfield  {journal} {\bibinfo  {journal} {Sov.
  Phys. JETP}\ }\textbf {\bibinfo {volume} {26}},\ \bibinfo {pages} {979}
  (\bibinfo {year} {1968})}\BibitemShut {NoStop}%
\bibitem [{\citenamefont {Korovin}(1971)}]{Korovin1971}%
  \BibitemOpen
  \bibfield  {author} {\bibinfo {author} {\bibfnamefont {L.~I.}\ \bibnamefont
  {Korovin}},\ }\href@noop {} {\bibfield  {journal} {\bibinfo  {journal} {Sov.
  Phys. Solid State}\ }\textbf {\bibinfo {volume} {13}},\ \bibinfo {pages}
  {695} (\bibinfo {year} {1971})}\BibitemShut {NoStop}%
\bibitem [{\citenamefont {Lang}\ \emph {et~al.}(2005)\citenamefont {Lang},
  \citenamefont {Korovin},\ and\ \citenamefont {Pavlov}}]{Lang2005}%
  \BibitemOpen
  \bibfield  {author} {\bibinfo {author} {\bibfnamefont {I.~G.}\ \bibnamefont
  {Lang}}, \bibinfo {author} {\bibfnamefont {L.~I.}\ \bibnamefont {Korovin}}, \
  and\ \bibinfo {author} {\bibfnamefont {S.~T.}\ \bibnamefont {Pavlov}},\
  }\href@noop {} {\bibfield  {journal} {\bibinfo  {journal} {Sov. Phys. Solid
  State}\ }\textbf {\bibinfo {volume} {47}},\ \bibinfo {pages} {1771} (\bibinfo
  {year} {2005})}\BibitemShut {NoStop}%
\bibitem [{\citenamefont {Novoselov}\ \emph {et~al.}(2007)\citenamefont
  {Novoselov}, \citenamefont {Jiang}, \citenamefont {Zhang}, \citenamefont
  {Morozov}, \citenamefont {Stormer}, \citenamefont {Zeitler}, \citenamefont
  {Maan}, \citenamefont {Boebinger}, \citenamefont {Kim},\ and\ \citenamefont
  {Geim}}]{Novoselov2007}%
  \BibitemOpen
  \bibfield  {author} {\bibinfo {author} {\bibfnamefont {K.~S.}\ \bibnamefont
  {Novoselov}}, \bibinfo {author} {\bibfnamefont {Z.}~\bibnamefont {Jiang}},
  \bibinfo {author} {\bibfnamefont {Y.}~\bibnamefont {Zhang}}, \bibinfo
  {author} {\bibfnamefont {S.~V.}\ \bibnamefont {Morozov}}, \bibinfo {author}
  {\bibfnamefont {H.~L.}\ \bibnamefont {Stormer}}, \bibinfo {author}
  {\bibfnamefont {U.}~\bibnamefont {Zeitler}}, \bibinfo {author} {\bibfnamefont
  {J.~C.}\ \bibnamefont {Maan}}, \bibinfo {author} {\bibfnamefont {G.~S.}\
  \bibnamefont {Boebinger}}, \bibinfo {author} {\bibfnamefont {P.}~\bibnamefont
  {Kim}}, \ and\ \bibinfo {author} {\bibfnamefont {A.~K.}\ \bibnamefont
  {Geim}},\ }\href {\doibase 10.1126/science.1137201} {\bibfield  {journal}
  {\bibinfo  {journal} {Science}\ }\textbf {\bibinfo {volume} {315}},\ \bibinfo
  {pages} {1379} (\bibinfo {year} {2007})}\BibitemShut {NoStop}%
\bibitem [{\citenamefont {Faugeras}\ \emph {et~al.}(2009)\citenamefont
  {Faugeras}, \citenamefont {Amado}, \citenamefont {Kossacki}, \citenamefont
  {Orlita}, \citenamefont {Sprinkle}, \citenamefont {Berger}, \citenamefont
  {de~Heer},\ and\ \citenamefont {Potemski}}]{Faugeras2009}%
  \BibitemOpen
  \bibfield  {author} {\bibinfo {author} {\bibfnamefont {C.}~\bibnamefont
  {Faugeras}}, \bibinfo {author} {\bibfnamefont {M.}~\bibnamefont {Amado}},
  \bibinfo {author} {\bibfnamefont {P.}~\bibnamefont {Kossacki}}, \bibinfo
  {author} {\bibfnamefont {M.}~\bibnamefont {Orlita}}, \bibinfo {author}
  {\bibfnamefont {M.}~\bibnamefont {Sprinkle}}, \bibinfo {author}
  {\bibfnamefont {C.}~\bibnamefont {Berger}}, \bibinfo {author} {\bibfnamefont
  {W.~A.}\ \bibnamefont {de~Heer}}, \ and\ \bibinfo {author} {\bibfnamefont
  {M.}~\bibnamefont {Potemski}},\ }\href {\doibase
  10.1103/PhysRevLett.103.186803} {\bibfield  {journal} {\bibinfo  {journal}
  {Phys. Rev. Lett.}\ }\textbf {\bibinfo {volume} {103}},\ \bibinfo {pages}
  {186803} (\bibinfo {year} {2009})}\BibitemShut {NoStop}%
\bibitem [{\citenamefont {Yan}\ \emph {et~al.}(2010)\citenamefont {Yan},
  \citenamefont {Goler}, \citenamefont {Rhone}, \citenamefont {Han},
  \citenamefont {He}, \citenamefont {Kim}, \citenamefont {Pellegrini},\ and\
  \citenamefont {Pinczuk}}]{Yan2010}%
  \BibitemOpen
  \bibfield  {author} {\bibinfo {author} {\bibfnamefont {J.}~\bibnamefont
  {Yan}}, \bibinfo {author} {\bibfnamefont {S.}~\bibnamefont {Goler}}, \bibinfo
  {author} {\bibfnamefont {T.~D.}\ \bibnamefont {Rhone}}, \bibinfo {author}
  {\bibfnamefont {M.}~\bibnamefont {Han}}, \bibinfo {author} {\bibfnamefont
  {R.}~\bibnamefont {He}}, \bibinfo {author} {\bibfnamefont {P.}~\bibnamefont
  {Kim}}, \bibinfo {author} {\bibfnamefont {V.}~\bibnamefont {Pellegrini}}, \
  and\ \bibinfo {author} {\bibfnamefont {A.}~\bibnamefont {Pinczuk}},\ }\href
  {\doibase 10.1103/PhysRevLett.105.227401} {\bibfield  {journal} {\bibinfo
  {journal} {Phys. Rev. Lett.}\ }\textbf {\bibinfo {volume} {105}},\ \bibinfo
  {pages} {227401} (\bibinfo {year} {2010})}\BibitemShut {NoStop}%
\bibitem [{\citenamefont {Ando}(2007)}]{Ando2007}%
  \BibitemOpen
  \bibfield  {author} {\bibinfo {author} {\bibfnamefont {T.}~\bibnamefont
  {Ando}},\ }\href {\doibase 10.1143/JPSJ.76.024712} {\bibfield  {journal}
  {\bibinfo  {journal} {J. Phys. Soc. Jpn.}\ }\textbf {\bibinfo {volume}
  {76}},\ \bibinfo {pages} {024712} (\bibinfo {year} {2007})}\BibitemShut
  {NoStop}%
\bibitem [{\citenamefont {Goerbig}\ \emph {et~al.}(2007)\citenamefont
  {Goerbig}, \citenamefont {Fuchs}, \citenamefont {Kechedzhi},\ and\
  \citenamefont {Fal'ko}}]{Goerbig2007}%
  \BibitemOpen
  \bibfield  {author} {\bibinfo {author} {\bibfnamefont {M.~O.}\ \bibnamefont
  {Goerbig}}, \bibinfo {author} {\bibfnamefont {J.-N.}\ \bibnamefont {Fuchs}},
  \bibinfo {author} {\bibfnamefont {K.}~\bibnamefont {Kechedzhi}}, \ and\
  \bibinfo {author} {\bibfnamefont {V.~I.}\ \bibnamefont {Fal'ko}},\
  }\href@noop {} {\bibfield  {journal} {\bibinfo  {journal} {Phys. Rev. Lett.}\
  }\textbf {\bibinfo {volume} {99}},\ \bibinfo {pages} {087402} (\bibinfo
  {year} {2007})}\BibitemShut {NoStop}%
\bibitem [{\citenamefont {Pound}\ \emph {et~al.}(2012)\citenamefont {Pound},
  \citenamefont {Carbotte},\ and\ \citenamefont {Nicol}}]{Pound2012}%
  \BibitemOpen
  \bibfield  {author} {\bibinfo {author} {\bibfnamefont {A.}~\bibnamefont
  {Pound}}, \bibinfo {author} {\bibfnamefont {J.~P.}\ \bibnamefont {Carbotte}},
  \ and\ \bibinfo {author} {\bibfnamefont {E.~J.}\ \bibnamefont {Nicol}},\
  }\href {\doibase 10.1103/PhysRevB.85.125422} {\bibfield  {journal} {\bibinfo
  {journal} {Phys. Rev. B}\ }\textbf {\bibinfo {volume} {85}},\ \bibinfo
  {pages} {125422} (\bibinfo {year} {2012})}\BibitemShut {NoStop}%
\bibitem [{\citenamefont {Li}\ \emph {et~al.}(2009)\citenamefont {Li},
  \citenamefont {Luican},\ and\ \citenamefont {Andrei}}]{Li2009}%
  \BibitemOpen
  \bibfield  {author} {\bibinfo {author} {\bibfnamefont {G.}~\bibnamefont
  {Li}}, \bibinfo {author} {\bibfnamefont {A.}~\bibnamefont {Luican}}, \ and\
  \bibinfo {author} {\bibfnamefont {E.~Y.}\ \bibnamefont {Andrei}},\ }\href
  {\doibase 10.1103/PhysRevLett.102.176804} {\bibfield  {journal} {\bibinfo
  {journal} {Phys. Rev. Lett.}\ }\textbf {\bibinfo {volume} {102}},\ \bibinfo
  {pages} {176804} (\bibinfo {year} {2009})}\BibitemShut {NoStop}%
\bibitem [{\citenamefont {Neugebauer}\ \emph {et~al.}(2009)\citenamefont
  {Neugebauer}, \citenamefont {Orlita}, \citenamefont {Faugeras}, \citenamefont
  {Barra},\ and\ \citenamefont {Potemski}}]{Neugebauer2009}%
  \BibitemOpen
  \bibfield  {author} {\bibinfo {author} {\bibfnamefont {P.}~\bibnamefont
  {Neugebauer}}, \bibinfo {author} {\bibfnamefont {M.}~\bibnamefont {Orlita}},
  \bibinfo {author} {\bibfnamefont {C.}~\bibnamefont {Faugeras}}, \bibinfo
  {author} {\bibfnamefont {A.-L.}\ \bibnamefont {Barra}}, \ and\ \bibinfo
  {author} {\bibfnamefont {M.}~\bibnamefont {Potemski}},\ }\href {\doibase
  10.1103/PhysRevLett.103.136403} {\bibfield  {journal} {\bibinfo  {journal}
  {Phys. Rev. Lett.}\ }\textbf {\bibinfo {volume} {103}},\ \bibinfo {pages}
  {136403} (\bibinfo {year} {2009})}\BibitemShut {NoStop}%
\bibitem [{\citenamefont {Faugeras}\ \emph {et~al.}(2011)\citenamefont
  {Faugeras}, \citenamefont {Amado}, \citenamefont {Kossacki}, \citenamefont
  {Orlita}, \citenamefont {K\"uhne}, \citenamefont {Nicolet}, \citenamefont
  {Latyshev},\ and\ \citenamefont {Potemski}}]{Faugeras2011}%
  \BibitemOpen
  \bibfield  {author} {\bibinfo {author} {\bibfnamefont {C.}~\bibnamefont
  {Faugeras}}, \bibinfo {author} {\bibfnamefont {M.}~\bibnamefont {Amado}},
  \bibinfo {author} {\bibfnamefont {P.}~\bibnamefont {Kossacki}}, \bibinfo
  {author} {\bibfnamefont {M.}~\bibnamefont {Orlita}}, \bibinfo {author}
  {\bibfnamefont {M.}~\bibnamefont {K\"uhne}}, \bibinfo {author} {\bibfnamefont
  {A.~A.~L.}\ \bibnamefont {Nicolet}}, \bibinfo {author} {\bibfnamefont
  {Y.~I.}\ \bibnamefont {Latyshev}}, \ and\ \bibinfo {author} {\bibfnamefont
  {M.}~\bibnamefont {Potemski}},\ }\href {\doibase
  10.1103/PhysRevLett.107.036807} {\bibfield  {journal} {\bibinfo  {journal}
  {Phys. Rev. Lett.}\ }\textbf {\bibinfo {volume} {107}},\ \bibinfo {pages}
  {036807} (\bibinfo {year} {2011})}\BibitemShut {NoStop}%
\bibitem [{\citenamefont {K\"uhne}\ \emph {et~al.}(2012)\citenamefont
  {K\"uhne}, \citenamefont {Faugeras}, \citenamefont {Kossacki}, \citenamefont
  {Nicolet}, \citenamefont {Orlita}, \citenamefont {Latyshev},\ and\
  \citenamefont {Potemski}}]{Kuhne2012}%
  \BibitemOpen
  \bibfield  {author} {\bibinfo {author} {\bibfnamefont {M.}~\bibnamefont
  {K\"uhne}}, \bibinfo {author} {\bibfnamefont {C.}~\bibnamefont {Faugeras}},
  \bibinfo {author} {\bibfnamefont {P.}~\bibnamefont {Kossacki}}, \bibinfo
  {author} {\bibfnamefont {A.~A.~L.}\ \bibnamefont {Nicolet}}, \bibinfo
  {author} {\bibfnamefont {M.}~\bibnamefont {Orlita}}, \bibinfo {author}
  {\bibfnamefont {Y.~I.}\ \bibnamefont {Latyshev}}, \ and\ \bibinfo {author}
  {\bibfnamefont {M.}~\bibnamefont {Potemski}},\ }\href {\doibase
  10.1103/PhysRevB.85.195406} {\bibfield  {journal} {\bibinfo  {journal} {Phys.
  Rev. B}\ }\textbf {\bibinfo {volume} {85}},\ \bibinfo {pages} {195406}
  (\bibinfo {year} {2012})}\BibitemShut {NoStop}%
\bibitem [{\citenamefont {Qiu}\ \emph {et~al.}(2013)\citenamefont {Qiu},
  \citenamefont {Shen}, \citenamefont {Cao}, \citenamefont {Cong},
  \citenamefont {Saito}, \citenamefont {Yu}, \citenamefont {Dresselhaus},\ and\
  \citenamefont {Yu}}]{Qiu2013}%
  \BibitemOpen
  \bibfield  {author} {\bibinfo {author} {\bibfnamefont {C.}~\bibnamefont
  {Qiu}}, \bibinfo {author} {\bibfnamefont {X.}~\bibnamefont {Shen}}, \bibinfo
  {author} {\bibfnamefont {B.}~\bibnamefont {Cao}}, \bibinfo {author}
  {\bibfnamefont {C.}~\bibnamefont {Cong}}, \bibinfo {author} {\bibfnamefont
  {R.}~\bibnamefont {Saito}}, \bibinfo {author} {\bibfnamefont
  {J.}~\bibnamefont {Yu}}, \bibinfo {author} {\bibfnamefont {M.~S.}\
  \bibnamefont {Dresselhaus}}, \ and\ \bibinfo {author} {\bibfnamefont
  {T.}~\bibnamefont {Yu}},\ }\href@noop {} {\bibfield  {journal} {\bibinfo
  {journal} {Phys. Rev. B}\ }\textbf {\bibinfo {volume} {88}},\ \bibinfo
  {pages} {165407} (\bibinfo {year} {2013})}\BibitemShut {NoStop}%
\bibitem [{\citenamefont {Ferrari}\ and\ \citenamefont
  {Basko}(2013)}]{Ferrari2013}%
  \BibitemOpen
  \bibfield  {author} {\bibinfo {author} {\bibfnamefont {A.~C.}\ \bibnamefont
  {Ferrari}}\ and\ \bibinfo {author} {\bibfnamefont {D.~M.}\ \bibnamefont
  {Basko}},\ }\href@noop {} {\bibfield  {journal} {\bibinfo  {journal} {Nature
  Nanotech.}\ }\textbf {\bibinfo {volume} {8}},\ \bibinfo {pages} {235}
  (\bibinfo {year} {2013})}\BibitemShut {NoStop}%
\bibitem [{\citenamefont {Berciaud}\ \emph {et~al.}(2014)\citenamefont
  {Berciaud}, \citenamefont {Potemski},\ and\ \citenamefont
  {Faugeras}}]{Berciaud2014}%
  \BibitemOpen
  \bibfield  {author} {\bibinfo {author} {\bibfnamefont {S.}~\bibnamefont
  {Berciaud}}, \bibinfo {author} {\bibfnamefont {M.}~\bibnamefont {Potemski}},
  \ and\ \bibinfo {author} {\bibfnamefont {C.}~\bibnamefont {Faugeras}},\
  }\href {\doibase 10.1021/nl501578m} {\bibfield  {journal} {\bibinfo
  {journal} {Nano Letters}\ }\textbf {\bibinfo {volume} {14}},\ \bibinfo
  {pages} {4548} (\bibinfo {year} {2014})}\BibitemShut {NoStop}%
\bibitem [{\citenamefont {Kashuba}\ and\ \citenamefont
  {Fal'ko}(2009)}]{Kashuba2009}%
  \BibitemOpen
  \bibfield  {author} {\bibinfo {author} {\bibfnamefont {O.}~\bibnamefont
  {Kashuba}}\ and\ \bibinfo {author} {\bibfnamefont {V.~I.}\ \bibnamefont
  {Fal'ko}},\ }\href {\doibase 10.1103/PhysRevB.80.241404} {\bibfield
  {journal} {\bibinfo  {journal} {Phys. Rev. B}\ }\textbf {\bibinfo {volume}
  {80}},\ \bibinfo {pages} {241404} (\bibinfo {year} {2009})}\BibitemShut
  {NoStop}%
\bibitem [{\citenamefont {Basko}(2008)}]{Basko2008}%
  \BibitemOpen
  \bibfield  {author} {\bibinfo {author} {\bibfnamefont {D.~M.}\ \bibnamefont
  {Basko}},\ }\href@noop {} {\bibfield  {journal} {\bibinfo  {journal} {Phys.
  Rev. B}\ }\textbf {\bibinfo {volume} {78}},\ \bibinfo {pages} {125418}
  (\bibinfo {year} {2008})}\BibitemShut {NoStop}%
\bibitem [{\citenamefont {Faugeras}\ \emph {et~al.}(2010)\citenamefont
  {Faugeras}, \citenamefont {Kossacki}, \citenamefont {Basko}, \citenamefont
  {Amado}, \citenamefont {Sprinkle}, \citenamefont {Berger}, \citenamefont
  {de~Heer},\ and\ \citenamefont {Potemski}}]{Faugeras2010a}%
  \BibitemOpen
  \bibfield  {author} {\bibinfo {author} {\bibfnamefont {C.}~\bibnamefont
  {Faugeras}}, \bibinfo {author} {\bibfnamefont {P.}~\bibnamefont {Kossacki}},
  \bibinfo {author} {\bibfnamefont {D.~M.}\ \bibnamefont {Basko}}, \bibinfo
  {author} {\bibfnamefont {M.}~\bibnamefont {Amado}}, \bibinfo {author}
  {\bibfnamefont {M.}~\bibnamefont {Sprinkle}}, \bibinfo {author}
  {\bibfnamefont {C.}~\bibnamefont {Berger}}, \bibinfo {author} {\bibfnamefont
  {W.~A.}\ \bibnamefont {de~Heer}}, \ and\ \bibinfo {author} {\bibfnamefont
  {M.}~\bibnamefont {Potemski}},\ }\href {\doibase 10.1103/PhysRevB.81.155436}
  {\bibfield  {journal} {\bibinfo  {journal} {Phys. Rev. B}\ }\textbf {\bibinfo
  {volume} {81}},\ \bibinfo {pages} {155436} (\bibinfo {year}
  {2010})}\BibitemShut {NoStop}%
\bibitem [{\citenamefont {Venezuela}\ \emph {et~al.}(2011)\citenamefont
  {Venezuela}, \citenamefont {Lazzeri},\ and\ \citenamefont
  {Mauri}}]{Venezuela2011}%
  \BibitemOpen
  \bibfield  {author} {\bibinfo {author} {\bibfnamefont {P.}~\bibnamefont
  {Venezuela}}, \bibinfo {author} {\bibfnamefont {M.}~\bibnamefont {Lazzeri}},
  \ and\ \bibinfo {author} {\bibfnamefont {F.}~\bibnamefont {Mauri}},\ }\href
  {\doibase 10.1103/PhysRevB.84.035433} {\bibfield  {journal} {\bibinfo
  {journal} {Phys. Rev. B}\ }\textbf {\bibinfo {volume} {84}},\ \bibinfo
  {pages} {035433} (\bibinfo {year} {2011})}\BibitemShut {NoStop}%
\bibitem [{\citenamefont {Piscanec}\ \emph {et~al.}(2004)\citenamefont
  {Piscanec}, \citenamefont {Lazzeri}, \citenamefont {Mauri}, \citenamefont
  {Ferrari},\ and\ \citenamefont {Robertson}}]{Piscanec2004}%
  \BibitemOpen
  \bibfield  {author} {\bibinfo {author} {\bibfnamefont {S.}~\bibnamefont
  {Piscanec}}, \bibinfo {author} {\bibfnamefont {M.}~\bibnamefont {Lazzeri}},
  \bibinfo {author} {\bibfnamefont {F.}~\bibnamefont {Mauri}}, \bibinfo
  {author} {\bibfnamefont {A.~C.}\ \bibnamefont {Ferrari}}, \ and\ \bibinfo
  {author} {\bibfnamefont {J.}~\bibnamefont {Robertson}},\ }\href {\doibase
  10.1103/PhysRevLett.93.185503} {\bibfield  {journal} {\bibinfo  {journal}
  {Phys. Rev. Lett.}\ }\textbf {\bibinfo {volume} {93}},\ \bibinfo {pages}
  {185503} (\bibinfo {year} {2004})}\BibitemShut {NoStop}%
\bibitem [{\citenamefont {Gr\"uneis}\ \emph {et~al.}(2009)\citenamefont
  {Gr\"uneis}, \citenamefont {Serrano}, \citenamefont {Bosak}, \citenamefont
  {Lazzeri}, \citenamefont {Molodtsov}, \citenamefont {Wirtz}, \citenamefont
  {Attaccalite}, \citenamefont {Krisch}, \citenamefont {Rubio}, \citenamefont
  {Mauri},\ and\ \citenamefont {Pichler}}]{Gruneis2009b}%
  \BibitemOpen
  \bibfield  {author} {\bibinfo {author} {\bibfnamefont {A.}~\bibnamefont
  {Gr\"uneis}}, \bibinfo {author} {\bibfnamefont {J.}~\bibnamefont {Serrano}},
  \bibinfo {author} {\bibfnamefont {A.}~\bibnamefont {Bosak}}, \bibinfo
  {author} {\bibfnamefont {M.}~\bibnamefont {Lazzeri}}, \bibinfo {author}
  {\bibfnamefont {S.~L.}\ \bibnamefont {Molodtsov}}, \bibinfo {author}
  {\bibfnamefont {L.}~\bibnamefont {Wirtz}}, \bibinfo {author} {\bibfnamefont
  {C.}~\bibnamefont {Attaccalite}}, \bibinfo {author} {\bibfnamefont
  {M.}~\bibnamefont {Krisch}}, \bibinfo {author} {\bibfnamefont
  {A.}~\bibnamefont {Rubio}}, \bibinfo {author} {\bibfnamefont
  {F.}~\bibnamefont {Mauri}}, \ and\ \bibinfo {author} {\bibfnamefont
  {T.}~\bibnamefont {Pichler}},\ }\href {\doibase 10.1103/PhysRevB.80.085423}
  {\bibfield  {journal} {\bibinfo  {journal} {Phys. Rev. B}\ }\textbf {\bibinfo
  {volume} {80}},\ \bibinfo {pages} {085423} (\bibinfo {year}
  {2009})}\BibitemShut {NoStop}%
\bibitem [{\citenamefont {Orlita}\ \emph {et~al.}(2012)\citenamefont {Orlita},
  \citenamefont {Tan}, \citenamefont {Potemski}, \citenamefont {Sprinkle},
  \citenamefont {Berger}, \citenamefont {de~Heer}, \citenamefont {Louie},\ and\
  \citenamefont {Martinez}}]{Orlita2012}%
  \BibitemOpen
  \bibfield  {author} {\bibinfo {author} {\bibfnamefont {M.}~\bibnamefont
  {Orlita}}, \bibinfo {author} {\bibfnamefont {L.~Z.}\ \bibnamefont {Tan}},
  \bibinfo {author} {\bibfnamefont {M.}~\bibnamefont {Potemski}}, \bibinfo
  {author} {\bibfnamefont {M.}~\bibnamefont {Sprinkle}}, \bibinfo {author}
  {\bibfnamefont {C.}~\bibnamefont {Berger}}, \bibinfo {author} {\bibfnamefont
  {W.~A.}\ \bibnamefont {de~Heer}}, \bibinfo {author} {\bibfnamefont {S.~G.}\
  \bibnamefont {Louie}}, \ and\ \bibinfo {author} {\bibfnamefont
  {G.}~\bibnamefont {Martinez}},\ }\href {\doibase
  10.1103/PhysRevLett.108.247401} {\bibfield  {journal} {\bibinfo  {journal}
  {Phys. Rev. Lett.}\ }\textbf {\bibinfo {volume} {108}},\ \bibinfo {pages}
  {247401} (\bibinfo {year} {2012})}\BibitemShut {NoStop}%
\bibitem [{\citenamefont {Kossacki}\ \emph {et~al.}(2012)\citenamefont
  {Kossacki}, \citenamefont {Faugeras}, \citenamefont {K\"uhne}, \citenamefont
  {Orlita}, \citenamefont {Mahmood}, \citenamefont {Dujardin}, \citenamefont
  {Nair}, \citenamefont {Geim},\ and\ \citenamefont {Potemski}}]{Kossacki2012}%
  \BibitemOpen
  \bibfield  {author} {\bibinfo {author} {\bibfnamefont {P.}~\bibnamefont
  {Kossacki}}, \bibinfo {author} {\bibfnamefont {C.}~\bibnamefont {Faugeras}},
  \bibinfo {author} {\bibfnamefont {M.}~\bibnamefont {K\"uhne}}, \bibinfo
  {author} {\bibfnamefont {M.}~\bibnamefont {Orlita}}, \bibinfo {author}
  {\bibfnamefont {A.}~\bibnamefont {Mahmood}}, \bibinfo {author} {\bibfnamefont
  {E.}~\bibnamefont {Dujardin}}, \bibinfo {author} {\bibfnamefont {R.~R.}\
  \bibnamefont {Nair}}, \bibinfo {author} {\bibfnamefont {A.~K.}\ \bibnamefont
  {Geim}}, \ and\ \bibinfo {author} {\bibfnamefont {M.}~\bibnamefont
  {Potemski}},\ }\href {\doibase 10.1103/PhysRevB.86.205431} {\bibfield
  {journal} {\bibinfo  {journal} {Phys. Rev. B}\ }\textbf {\bibinfo {volume}
  {86}},\ \bibinfo {pages} {205431} (\bibinfo {year} {2012})}\BibitemShut
  {NoStop}%
\bibitem [{\citenamefont {Faugeras}\ \emph {et~al.}(2014)\citenamefont
  {Faugeras}, \citenamefont {Binder}, \citenamefont {Nicolet}, \citenamefont
  {Leszczynski}, \citenamefont {Kossacki}, \citenamefont {Wysmolek},
  \citenamefont {Orlita},\ and\ \citenamefont {Potemski}}]{Faugeras2014}%
  \BibitemOpen
  \bibfield  {author} {\bibinfo {author} {\bibfnamefont {C.}~\bibnamefont
  {Faugeras}}, \bibinfo {author} {\bibfnamefont {J.}~\bibnamefont {Binder}},
  \bibinfo {author} {\bibfnamefont {A.~A.~L.}\ \bibnamefont {Nicolet}},
  \bibinfo {author} {\bibfnamefont {P.}~\bibnamefont {Leszczynski}}, \bibinfo
  {author} {\bibfnamefont {P.}~\bibnamefont {Kossacki}}, \bibinfo {author}
  {\bibfnamefont {A.}~\bibnamefont {Wysmolek}}, \bibinfo {author}
  {\bibfnamefont {M.}~\bibnamefont {Orlita}}, \ and\ \bibinfo {author}
  {\bibfnamefont {M.}~\bibnamefont {Potemski}},\ }\href@noop {} {\bibfield
  {journal} {\bibinfo  {journal} {EuroPhys. Lett.}\ } (\bibinfo {year}
  {2014})}\BibitemShut {NoStop}%
\bibitem [{\citenamefont {Faugeras}\ \emph {et~al.}(2015)\citenamefont
  {Faugeras}, \citenamefont {Berciaud}, \citenamefont {Leszczynski},
  \citenamefont {Henni}, \citenamefont {Nogajewski}, \citenamefont {Orlita},
  \citenamefont {Taniguchi}, \citenamefont {Watanabe}, \citenamefont
  {Forsythe}, \citenamefont {Kim}, \citenamefont {Jalil}, \citenamefont {Geim},
  \citenamefont {Basko},\ and\ \citenamefont {Potemski}}]{Faugeras2015}%
  \BibitemOpen
  \bibfield  {author} {\bibinfo {author} {\bibfnamefont {C.}~\bibnamefont
  {Faugeras}}, \bibinfo {author} {\bibfnamefont {S.}~\bibnamefont {Berciaud}},
  \bibinfo {author} {\bibfnamefont {P.}~\bibnamefont {Leszczynski}}, \bibinfo
  {author} {\bibfnamefont {Y.}~\bibnamefont {Henni}}, \bibinfo {author}
  {\bibfnamefont {K.}~\bibnamefont {Nogajewski}}, \bibinfo {author}
  {\bibfnamefont {M.}~\bibnamefont {Orlita}}, \bibinfo {author} {\bibfnamefont
  {T.}~\bibnamefont {Taniguchi}}, \bibinfo {author} {\bibfnamefont
  {K.}~\bibnamefont {Watanabe}}, \bibinfo {author} {\bibfnamefont
  {C.}~\bibnamefont {Forsythe}}, \bibinfo {author} {\bibfnamefont
  {P.}~\bibnamefont {Kim}}, \bibinfo {author} {\bibfnamefont {R.}~\bibnamefont
  {Jalil}}, \bibinfo {author} {\bibfnamefont {A.~K.}\ \bibnamefont {Geim}},
  \bibinfo {author} {\bibfnamefont {D.~M.}\ \bibnamefont {Basko}}, \ and\
  \bibinfo {author} {\bibfnamefont {M.}~\bibnamefont {Potemski}},\ }\href
  {\doibase 10.1103/PhysRevLett.114.126804} {\bibfield  {journal} {\bibinfo
  {journal} {Phys. Rev. Lett.}\ }\textbf {\bibinfo {volume} {114}},\ \bibinfo
  {pages} {126804} (\bibinfo {year} {2015})}\BibitemShut {NoStop}%
\bibitem [{\citenamefont {Pisana}\ \emph {et~al.}(2007)\citenamefont {Pisana},
  \citenamefont {Lazzeri}, \citenamefont {Casiraghi}, \citenamefont
  {Novoselov}, \citenamefont {Geim}, \citenamefont {Ferrari},\ and\
  \citenamefont {Mauri}}]{Pisana2007}%
  \BibitemOpen
  \bibfield  {author} {\bibinfo {author} {\bibfnamefont {S.}~\bibnamefont
  {Pisana}}, \bibinfo {author} {\bibfnamefont {M.}~\bibnamefont {Lazzeri}},
  \bibinfo {author} {\bibfnamefont {C.}~\bibnamefont {Casiraghi}}, \bibinfo
  {author} {\bibfnamefont {K.~S.}\ \bibnamefont {Novoselov}}, \bibinfo {author}
  {\bibfnamefont {A.~K.}\ \bibnamefont {Geim}}, \bibinfo {author}
  {\bibfnamefont {A.~C.}\ \bibnamefont {Ferrari}}, \ and\ \bibinfo {author}
  {\bibfnamefont {F.}~\bibnamefont {Mauri}},\ }\href@noop {} {\bibfield
  {journal} {\bibinfo  {journal} {Nature Mat.}\ }\textbf {\bibinfo {volume}
  {6}},\ \bibinfo {pages} {198} (\bibinfo {year} {2007})}\BibitemShut {NoStop}%
\bibitem [{\citenamefont {Yan}\ \emph {et~al.}(2007)\citenamefont {Yan},
  \citenamefont {Zhang}, \citenamefont {Kim},\ and\ \citenamefont
  {Pinczuk}}]{Yan2007}%
  \BibitemOpen
  \bibfield  {author} {\bibinfo {author} {\bibfnamefont {J.}~\bibnamefont
  {Yan}}, \bibinfo {author} {\bibfnamefont {Y.}~\bibnamefont {Zhang}}, \bibinfo
  {author} {\bibfnamefont {P.}~\bibnamefont {Kim}}, \ and\ \bibinfo {author}
  {\bibfnamefont {A.}~\bibnamefont {Pinczuk}},\ }\href {\doibase
  10.1103/PhysRevLett.98.166802} {\bibfield  {journal} {\bibinfo  {journal}
  {Phys. Rev. Lett.}\ }\textbf {\bibinfo {volume} {98}},\ \bibinfo {pages}
  {166802} (\bibinfo {year} {2007})}\BibitemShut {NoStop}%
\bibitem [{\citenamefont {Basko}\ and\ \citenamefont
  {Aleiner}(2008)}]{BaskoAleiner2008}%
  \BibitemOpen
  \bibfield  {author} {\bibinfo {author} {\bibfnamefont {D.~M.}\ \bibnamefont
  {Basko}}\ and\ \bibinfo {author} {\bibfnamefont {I.~L.}\ \bibnamefont
  {Aleiner}},\ }\href {\doibase 10.1103/PhysRevB.77.041409} {\bibfield
  {journal} {\bibinfo  {journal} {Phys. Rev. B}\ }\textbf {\bibinfo {volume}
  {77}},\ \bibinfo {pages} {041409} (\bibinfo {year} {2008})}\BibitemShut
  {NoStop}%
\bibitem [{\citenamefont {Lazzeri}\ \emph {et~al.}(2008)\citenamefont
  {Lazzeri}, \citenamefont {Attaccalite}, \citenamefont {Wirtz},\ and\
  \citenamefont {Mauri}}]{Lazzeri2008}%
  \BibitemOpen
  \bibfield  {author} {\bibinfo {author} {\bibfnamefont {M.}~\bibnamefont
  {Lazzeri}}, \bibinfo {author} {\bibfnamefont {C.}~\bibnamefont
  {Attaccalite}}, \bibinfo {author} {\bibfnamefont {L.}~\bibnamefont {Wirtz}},
  \ and\ \bibinfo {author} {\bibfnamefont {F.}~\bibnamefont {Mauri}},\ }\href
  {\doibase 10.1103/PhysRevB.78.081406} {\bibfield  {journal} {\bibinfo
  {journal} {Phys. Rev. B}\ }\textbf {\bibinfo {volume} {78}},\ \bibinfo
  {pages} {081406} (\bibinfo {year} {2008})}\BibitemShut {NoStop}%
\bibitem [{\citenamefont {Abrikosov}\ \emph {et~al.}(1975)\citenamefont
  {Abrikosov}, \citenamefont {Gorkov},\ and\ \citenamefont
  {Dzyaloshinski}}]{AGD}%
  \BibitemOpen
  \bibfield  {author} {\bibinfo {author} {\bibfnamefont {A.~A.}\ \bibnamefont
  {Abrikosov}}, \bibinfo {author} {\bibfnamefont {L.~P.}\ \bibnamefont
  {Gorkov}}, \ and\ \bibinfo {author} {\bibfnamefont {I.~E.}\ \bibnamefont
  {Dzyaloshinski}},\ }\href@noop {} {\emph {\bibinfo {title} {Methods of
  Quantum Field Theory in Statistical Physics}}}\ (\bibinfo  {publisher}
  {Dover},\ \bibinfo {year} {1975})\BibitemShut {NoStop}%
\bibitem [{\citenamefont {Aryasetiawan}\ and\ \citenamefont
  {Gunnarsson}(1998)}]{Aryasetiawan1998}%
  \BibitemOpen
  \bibfield  {author} {\bibinfo {author} {\bibfnamefont {F.}~\bibnamefont
  {Aryasetiawan}}\ and\ \bibinfo {author} {\bibfnamefont {O.}~\bibnamefont
  {Gunnarsson}},\ }\href@noop {} {\bibfield  {journal} {\bibinfo  {journal}
  {Rep. Prog. Phys.}\ }\textbf {\bibinfo {volume} {61}},\ \bibinfo {pages}
  {237} (\bibinfo {year} {1998})}\BibitemShut {NoStop}%
\end{thebibliography}

%

\end{document}